\documentclass[article]{elsarticle}

\usepackage{lineno,hyperref}
\usepackage{multirow}
\usepackage{tabularx}
\usepackage{longtable}      
\usepackage[dvipsnames]{xcolor} 
\usepackage[dvipsnames]{xcolor}
\usepackage[utf8]{inputenc}
\usepackage[english]{babel}
\usepackage[T1]{fontenc}
\usepackage{amssymb}
\usepackage{multirow}
\usepackage{longtable}
\usepackage{graphicx}
\graphicspath{ {./images/} }
\usepackage{caption}
\usepackage{subcaption}
\usepackage{gensymb}
\usepackage{pdfpages}
\usepackage{amsmath}

\DeclareUnicodeCharacter{2260}{\neq} 
\DeclareUnicodeCharacter{2212}{-}   
\DeclareUnicodeCharacter{E5F8}{--}  

\DeclareUnicodeCharacter{2212}{-} 

\makeatletter
\def\ps@pprintTitle{%
  \let\@oddhead\@empty
  \let\@evenhead\@empty
  \def\@oddfoot{\reset@font\hfil\thepage\hfil}
  \let\@evenfoot\@oddfoot
}
\makeatother

\begin{document}

\begin{frontmatter}
\title{Interactive Multiscale Modeling to Bridge Atomic Properties and Electrochemical Performance in Li-CO$_2$ Battery Design}
\author{Mohammed Lemaalem $^{a,b}$, Selva Chandrasekaran Selvaraj $^{a,b}$, Ilias Papailias $^{c}$, Naveen K. Dandu $^{a,b}$, Arash Namaeighasemi $^{a}$, Larry A. Curtiss $^{b,*}$, Amin Salehi-Khojin $^{c,*}$, and Anh T. Ngo $^{a,b,*}$}
\address{$^{a}$ Department of Chemical Engineering, University of Illinois Chicago, Chicago, IL60608, USA} 
\address{$^{b}$ Materials Science Division, Argonne National Laboratory, Lemont, IL 60439, USA. }
\address{$^{c}$ Department of Mechanical Engineering, Lyle School of Engineering, SMU, 3101 Dyer Street Dallas, TX 75205, USA} 
\cortext[]{anhngo@uic.edu, asalehikhojin@mail.smu.edu, curtiss@anl.gov} 
\begin{abstract}
Li-CO$_2$ batteries are promising energy storage systems due to their high theoretical energy density and CO$_2$ fixation capability, relying on reversible Li$_2$CO$_3$/C formation during discharge/charge cycles. We present a multiscale modeling framework integrating Density Functional Theory (DFT), Ab-Initio Molecular Dynamics (AIMD), classical Molecular Dynamics (MD), and Finite Element Analysis (FEA) to investigate atomic and cell-level properties. The considered Li-CO$_2$ battery consists of a lithium metal anode, an ionic liquid electrolyte, and a carbon cloth cathode with Sb$_{0.67}$Bi$_{1.33}$Te$_3$ catalyst. DFT and AIMD determined the electrical conductivities of Sb$_{0.67}$Bi$_{1.33}$Te$_3$ and Li$_2$CO$_3$ using the Kubo-Greenwood formalism and studied the CO$_2$ reduction mechanism on the cathode catalyst. MD simulations calculated the CO$_2$ diffusion coefficient, Li$^+$ transference number, ionic conductivity, and Li$^+$ solvation structure. The FEA model, parameterized with atomistic simulations data, reproduced the available experimental voltage-capacity profile at 1 mA/cm$^2$ and revealed spatio-temporal variations in Li$_2$CO$_3$/C deposition, porosity, and CO$_2$ concentration dependence on discharge rates in the cathode. Accordingly, Li$_2$CO$_3$ can form large and thin film deposits, leading to dispersed and local porosity changes at 0.1 mA/cm$^2$ and 1 mA/cm$^2$, respectively. The capacity decreases exponentially from 81,570 mAh/g at 0.1 mA/cm$^2$ to 6,200 mAh/g at 1 mA/cm$^2$, due to pore clogging from excessive discharge product deposition that limits CO$_2$ transport to the cathode interior. Therefore, the performance of Li-CO$_2$ batteries can be improved by enhancing CO$_2$ transport, regulating Li$_2$CO$_3$ deposition, and optimizing cathode architecture.
\end{abstract}
\end{frontmatter}
\section{Introduction}
Lithium-carbon dioxide (Li-CO$_{2}$) batteries, with their high theoretical energy density and CO$_{2}$ fixation capability, have emerged as a potential energy storage solution \cite{lv2022metal, qiao2017li}. These batteries provide a theoretical specific energy density of 1876 Wh kg$^{-1}$, which is substantially higher than that of current lithium-ion battery systems \cite{ahmadiparidari2019long, jaradat2024fast}. This high energy density makes Li-CO$_{2}$ batteries particularly attractive for applications in aviation, aerospace, and electric vehicles, where lighter and more compact energy storage solutions are crucial \cite{pathak2023lithium}. The potential for electric vehicles with much longer ranges or considerably lighter batteries could revolutionize the automotive market \cite{cano2018batteries, pathak2023lithium}. However, Li-CO$_{2}$ batteries are still in early development stages and face challenges including reaction reversibility issues, low energy efficiency, and the need for improved cathode designs \cite{ezeigwe2022review}. Ongoing research and development efforts focus on overcoming these limitations, with recent advancements showing promising results \cite{jaradat2024fast, pan2024approaching, zhang2024unravelling, yao2024high, mu2024toward, liu2024deciphering, xiao2024recent, sandhiya2024compositionally}. The Li-CO$_{2}$ batteries operate on a unique electrochemical principle involving the reversible reaction between lithium and carbon dioxide \cite{lv2022metal, qiao2017li} contributing to CO$_{2}$ capture and utilization efforts \cite{mu2024toward, gittleson2014operando}. During the discharge process, lithium oxidizes at the anode, releasing lithium ions and electrons $(Li \rightarrow Li\textsuperscript{+} + e\textsuperscript{-})$. Besides, at the cathode in the presence of solid catalysts \cite{sun2023binuclear, jaradat2024fast}, CO\textsubscript{2} undergoes reduction ($3\text{CO}_2 + 4\text{e}^- \rightarrow 2\text{CO}_3^{2-} + \text{C}$), combining with lithium ions and electrons to form primarily lithium carbonate (Li$_{2}$CO$_{3}$) and carbon (C) $(4Li\textsuperscript{+} + 3CO\textsubscript{2} + 4e\textsuperscript{-} \rightarrow 2Li\textsubscript{2}CO\textsubscript{3} + C)$. This discharge reaction typically occurs at a voltage of approximately 2.8 V \cite{wang2023developing}. The charging process reverses these reactions, decomposing the discharge products to regenerate lithium and CO\textsubscript{2}. However, due to the stability of Li\textsubscript{2}CO\textsubscript{3}, this process requires higher voltages, often exceeding 3.8 V \cite{zhang2022challenges}. This increased voltage requirement underscores one of several key challenges facing Li-CO\textsubscript{2} battery technology. In particular, the formation of solid Li\textsubscript{2}CO\textsubscript{3} and carbon can lead to electrode clogging and increased internal resistance \cite{zhang2024unravelling, sarkar2024electrochemical}. Moreover, the high charging voltages needed to decompose Li\textsubscript{2}CO\textsubscript{3} significantly reduce the overall energy efficiency of the system \cite{ma2022unraveling, zhou2021fast}. To address these issues, effective cathode catalysts are critical for enhancing reaction kinetics and minimizing overpotentials in Li-CO\textsubscript{2} batteries \cite{pan2024approaching, jiao2021recent}. Nevertheless, unwanted side reactions, particularly electrolyte decomposition, can occur at these high voltages \cite{ezeigwe2022review}. Furthermore, the inherent inertness of CO\textsubscript{2} compounds the challenge, as it makes electrochemical reduction difficult and adversely affects both discharge voltage and energy efficiency \cite{yu2020recent, qiao2017li, gittleson2014operando}.\\
\begin{figure*}[!h]
	\begin{center}
		\begin{minipage}[t]{1\textwidth}
			\includegraphics[width=1\textwidth]{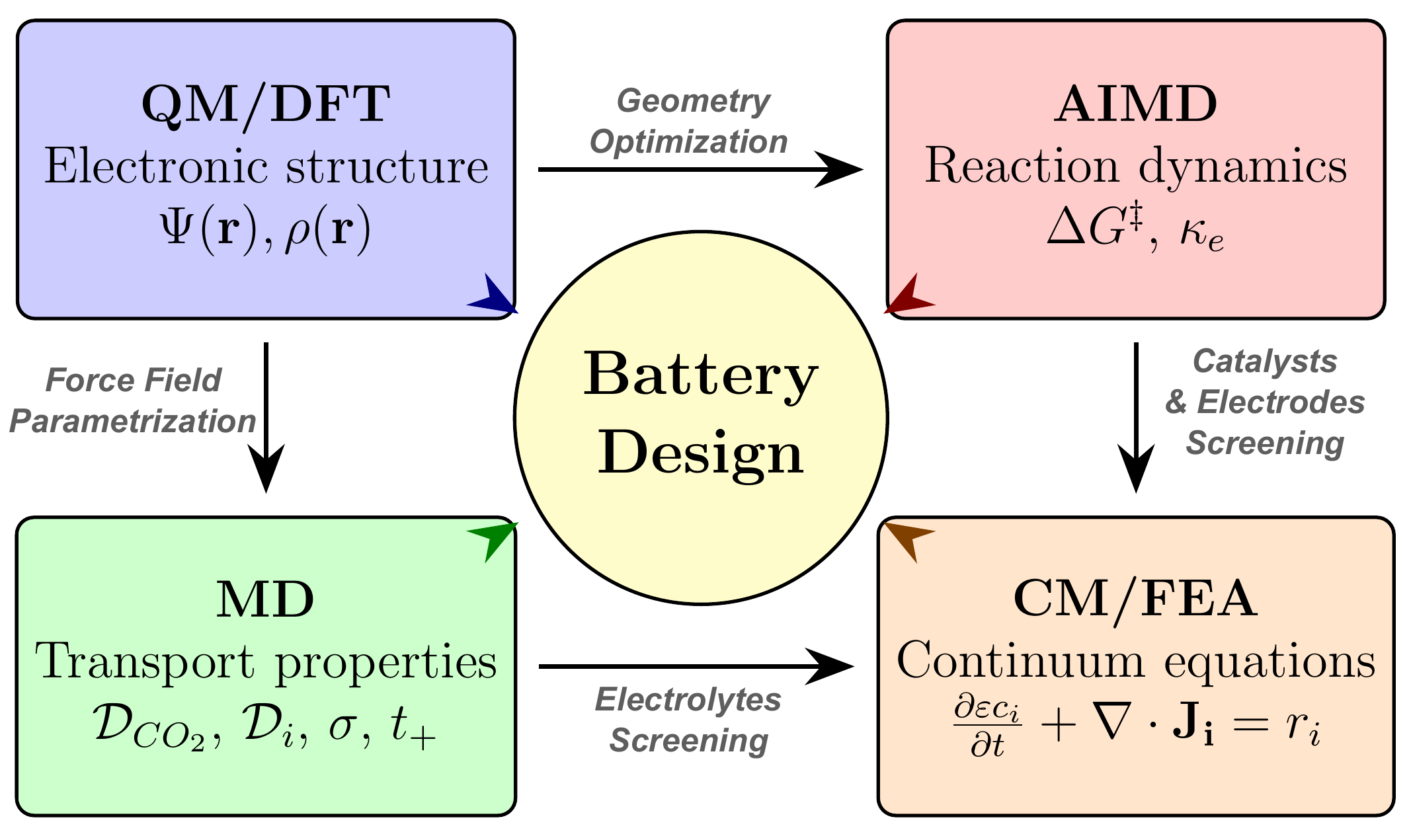}
            \caption{Diagram outlining the multiscale computational workflow for battery design, integrating four key methods: (1) quantum mechanics and Density Functional Theory calculations (QM/DFT), which calculate electronic structure and optimize atomic geometries; (2) Ab Initio Molecular Dynamics (AIMD), which simulates reaction dynamics to screen electrodes and catalysts; (3) Molecular Dynamics (MD), which models transport properties for electrolyte design; and (4) continuum modeling/Finite Element Analysis (CM/FEA), which develops and solves continuum equations to predict system-level performance. All methods feed into battery design, with arrows indicating critical data exchange.}
            \label{Figure1} 
		\end{minipage}
	\end{center}
\end{figure*}

In this work, we consider a Li-CO$_2$ battery cell comprising a lithium anode, a hybrid ionic liquid electrolyte, and a porous cathode. The electrolyte features a unique composition of 1-ethyl-3-methylimidazolium tetrafluoroborate (EMIM-BF$_4$) and dimethyl sulfoxide (DMSO) in a 2:3 ratio, supplemented with zinc iodide (ZnI$_2$) and 1 M lithium bis(trifluoromethanesulfonyl)imide (LiTFSI) salt. This optimized electrolyte formulation, experimentally validated for superior performance \cite{jaradat2024fast}, enables efficient ion transport and stability. The cathode utilizes carbon cloth coated with a novel transition metal trichalcogenide alloy catalyst (Sb$_{0.67}$Bi$_{1.33}$Te$_3$), which significantly enhances CO$_2$ reduction/oxidation kinetics and cycling stability \cite{jaradat2024fast}.
While recent advancements in Li-CO$_2$ batteries have focused on developing novel cathode materials and electrolytes \cite{yao2024high, shi2022highly, guo2023intrinsic, mu2024toward, sandhiya2024compositionally, xiao2024recent, xiao2020investigation}, most computational studies have been limited to atomistic-level investigations. To address this gap, we introduce the first comprehensive multiscale framework that integrates quantum-scale simulations, Molecular Dynamics, and continuum modeling of Li-CO$_2$ batteries. This approach uniquely connects atomic-scale phenomena, such as catalyst surface reactions and ion transport, to macroscopic cell performance, enabling holistic optimization of Li-CO$_2$ battery systems. This strategy is also needed for other battery types \cite{yu2024understanding, rizvi2024multiscale, yang2024multiscale, li2020multiscale}. Our methodology advances the field by addressing critical knowledge gaps in the spatiotemporal evolution of parameters and the interdependencies between electrochemical processes at different scales, and it can be extended to the study of other battery systems.

At the Li-CO$_2$ battery cell level, Finite Element Analysis (FEA) can effectively simulate charge/discharge cycles, enabling macroscopic analysis of complex processes within the battery cell. These processes include ion transport through porous electrodes, reaction kinetics, and their impact on cell performance \cite{bao2015discharge, mehta2021li, mukouyama2023finite}. However, FEA requires accurate electrode and electrolyte properties as input parameters, which can be derived from atomistic simulations.
To bridge the gap between atomic and cell-level simulations and to calculate the necessary parameters for FEA, we employed a closed-loop workflow from the atomic to the cell level, as summarized in Figure \ref{Figure1}. We employ quantum mechanics (QM) and Density Functional Theory (DFT) calculations to determine partial charges, equilibrated geometries, and Molecular Dynamics Force Field (FF) parameters for battery constituents using small systems (number of atoms < 1000). Ab Initio Molecular Dynamics (AIMD) calculations are used to determine the electrical conductivity of Li$_2$CO$_3$ and Sb$_{0.67}$Bi$_{1.33}$Te$_3$, based on DFT-equilibrated geometries and employing the Kubo-Greenwood formula \cite{7knyazev2013ab, 8kubo1957j, 9greenwood1958proc}. AIMD is also used to study the surface reactions of Sb$_{0.67}$Bi$_{1.33}$Te$_3$ as a cathode catalyst material. Large-scale MD simulations are conducted to investigate the transport properties of the liquid electrolyte. Specifically, we calculate the diffusion coefficients of Li$^+$ and CO$_2$, as well as the Li$^+$ transference number, by examining the mean square displacement (MSD) of the particles over time using the Einstein equation \cite{lemaalem2023effects}, and the ionic conductivity using the Green-Kubo formula \cite{tovey2023mdsuite}. The atomic-scale insights gained from DFT/AIMD and MD simulations are then integrated into our FEA model, enabling a comprehensive analysis of cell-level performance. Our FEA simulations investigate various aspects of battery cell behavior, including specific capacity, discharge behavior, and voltage-capacity profiles, considering factors such as CO$_2$ transport properties, cathode porosity, and the deposition of Li$_2$CO$_3$ and C.\\

The Li-CO$_2$ continuum model (CM) of the Li-CO$_2$ battery that we developed, supported by accurately calculated parameters from atomistic-level simulations, enables us to make predictions about porosity, deposited species quantities, and CO$_2$ concentration variations—parameters that are difficult to assess through experimental methods alone. Furthermore, it allows us to evaluate these properties as a function of time, providing dynamic insights into the battery's behavior at the cell level. This hierarchical, cross-validated workflow not only bridges the longstanding gap between atomistic understanding and macroscopic battery behavior but also enables targeted material screening and iterative optimization of cell components. The resulting predictive capability accelerates the discovery and engineering of Li-CO$_2$ batteries with enhanced energy density, cycle life, and operational safety. Moreover, our methodology provides a generalizable template for the rational design of other advanced battery chemistries, where multiscale interactions govern performance and reliability. Ultimately, this multiscale computational strategy offers a powerful tool to guide experimental efforts, reduce development time, and unlock the full potential of emerging energy storage technologies.
\section{Methods and computational details} \label{Methods}
\subsection{Density Functional Theory Calculations for Structure Optimization}
First-principles Density Functional Theory (DFT) calculations were performed using the Vienna Ab-initio Simulation Package (VASP) \cite{1kresse1993ab, 2kresse1994ab} to optimize the crystal structures of Sb$_{0.67}$Bi$_{1.33}$Te$_3$ and Li$_2$CO$_3$. The Sb$_{0.67}$Bi$_{1.33}$Te$_3$ structure was constructed based on Sb$_2$Te$_3$ from the Materials Project database \cite{jain2013commentary}, and multiple configurations were optimized to identify the most energetically favorable structure. The Li$_2$CO$_3$ structure was also obtained from the Materials Project database. For surface interaction studies, the Perdew-Burke-Ernzerhof (PBE) functional within the generalized gradient approximation (GGA) was used, and dispersion interactions were accounted for using the D3 van der Waals correction \cite{grimme2016dispersion}. These calculations are crucial for accurately modeling CO$_2$ adsorption and reduction processes on the catalyst surface. Further computational details are provided in the Supplementary Information.

\subsection{Ab Initio Molecular Dynamics (AIMD) and Kubo-Greenwood method for  electrical conductivity calculation} 
 Using optimized structures of Sb$_{0.67}$Bi$_{1.33}$Te$_3$ and Li$_2$CO$_3$ compounds from our DFT calculation, we carried out the Ab Initio Molecular Dynamics (AIMD) simulations at 300K up to 1ps with NVT ensemble and 1fs time step to select 10 different frames in the 1ps trajectory data. Subsequently, we generated wavefunctions ($\Psi_{i,k}$ where $i$ and $k$ are band and k-point indices, respectively) data for the selected frames using DFT calculations as explained by Knyazev et. al. \cite{6knyazev2019thermodynamic, 7knyazev2013ab} to compute the electronic conductivity based on Kubo-Greenwood (KG) formula. During the calculation of wavefunction data from VASP, we included 100\% more unoccupied bands than occupied bands for both Sb$_{0.67}$Bi$_{1.33}$Te$_3$ and Li$_2$CO$_3$ compounds to account for excited state wavefunctions and their corresponding energies. The dynamic electrical conductivity $\kappa(\omega)=\kappa_1(\omega)+i\kappa_2(\omega)$ is represented as a coefficient of the current density $\mathbf{J}(\omega)$ and electric field $\mathbf{E}(\omega)$:
\begin{equation}
\mathbf{J}(\omega)=[\kappa_1(\omega)+i\kappa_2(\omega)]\mathbf{E}(\omega)
\end{equation}
where $\omega$ is the frequency of the external electric field $\mathbf{E}(\omega)$. The important component $\kappa_1(\omega)$ relates to the energy absorbed by the electrons and is computed by the Kubo-Greenwood formula as \cite{7knyazev2013ab, 8kubo1957j, 9greenwood1958proc}
\begin{equation}
    \kappa_1 (\omega) = \frac{2\pi e^2 \hbar^2}{3m\omega\Omega} \sum_{i,j,\alpha,k} W(k) |\langle \Psi_{i,k} |\nabla_\alpha| \Psi_{j,k} \rangle|^2 [f(\varepsilon_{i,k}) - f(\varepsilon_{j,k})] \delta(\varepsilon_{j,k} - \varepsilon_{i,k} - \hbar\omega)
\end{equation}
Here, the sum is over all $\mathbf{k}$ points in the Brillouin zone, band indices $i$ and $j$ and three spatial dimensions $\alpha$. $\kappa_1(\omega)$ is the frequency-dependent electrical conductivity. $e$ is the elementary charge. $\hbar$ is the reduced Planck constant. $m$ is the electron mass. $\omega$ is the angular frequency. $\Omega$ is the volume of the system. $f_i$ and $f_j$ are the Fermi-Dirac distribution functions for states $i$ and $j$. $|\psi_i\rangle$ and $|\psi_j\rangle$ are the wavefunctions of states $i$ and $j$. $\varepsilon_{i,k}$ and $\varepsilon_{i,k} $ i are the energies of states $i$ and $j$. $\delta$ is the Dirac delta function. $W(k)$ denotes a weight function at a particular $\mathbf{k}$ point. The wave function $\Psi_{i,\mathbf{k}}$ is generated from our AIMD simulations at room temperature as explained above. $f(\varepsilon_{i,\mathbf{k}})$ is the Fermi function. In practical calculations, $\delta$-function is written in terms of the Gaussian function:

\begin{equation}
\delta(\varepsilon_{j,k} - \varepsilon_{i,k} - \hbar\omega) \rightarrow \frac{1}{\sqrt{2\pi}\Delta E} \exp\left(-\frac{(\varepsilon_{j,\mathbf{k}} - \varepsilon_{i,\mathbf{k}})}{2(\Delta E)^2}\right)
\end{equation}

where $\Delta$E is broadening of the $\delta$-function. This technical parameter plays an essential role in controlling the algorithm in the Kubo-Greenwood formula, particularly at low frequencies.

\subsection{Atomistic Molecular Dynamics Simulations (MD)} \label{MD}
Classical Molecular Dynamics (MD) simulations were performed using LAMMPS \cite{thompson2022lammps} to investigate the electrolyte system relevant to Li-CO$_2$ batteries. The simulated electrolyte consisted of DMSO:(EMIM$^{+}$, BF${4}^{-}$) at a 3:2 ratio, 1 M (Li$^{+}$, TFSI$^{-}$), and included 6000 DMSO molecules, 1857 (EMIM$^{+}$, BF${4}^{-}$) pairs, 566 (Li$^{+}$, TFSI$^{-}$), 14 ZnI$_2$, and 14 CO$_2$ molecules, totaling 113,142 atoms. Initial configurations were generated and energy minimized, followed by agitation using the Langevin and Nosé-Hoover thermostats. The system was equilibrated to ambient conditions (300 K, 1 bar) in stages using the Berendsen barostat, and transport properties were evaluated under experimental-like conditions in the NVT ensemble at 300 K. Further details on force field parameterization, cutoff schemes, and equilibration protocols are provided in the Supplementary Information.
\subsection{Finite element model}\label{section2.4}
\begin{figure*}[!h]
	\begin{center}
		\begin{minipage}[t]{1\textwidth}
			\includegraphics[width=1\textwidth]{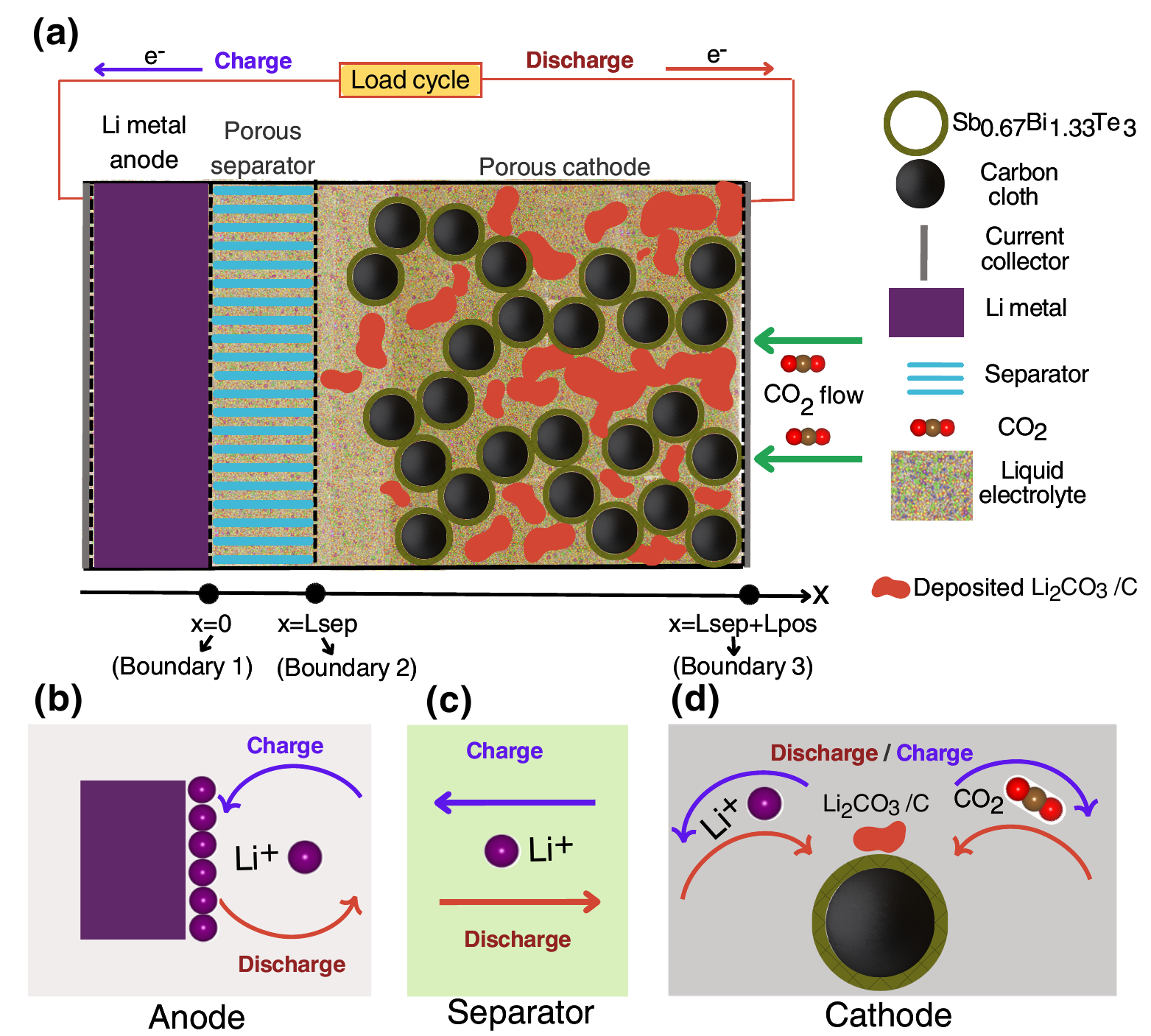}
            \caption{(a) Schematic diagram of the Li-CO$_2$ continuum model used in this study. The model consists of two active domains: the separator (from boundary 1 to boundary 2) and the porous electrode (from boundary 2 to boundary 3). The Li anode and CO$_2$ inlet are represented as boundaries 1 and 3, respectively. The inset at the bottom illustrates the key processes: (b) surface reactions at the anode, (c) ion transport within the separator, and (d) electrochemical reactions at the cathode.}
            \label{Figure2} 
		\end{minipage}
	\end{center}
\end{figure*}

A mathematical framework has been developed for modeling the behavior of prismatic Li-CO$_2$ batteries, drawing inspiration from previous studies on Li-O$_2$ and Li-ion batteries \cite{FEA1, FEA2, FEA3}. The model integrates key principles such as mass and current conservation, species transport phenomena, and reaction kinetics within the cathode and separator regions to provide deeper insights into the complex mechanisms that occur within the cell during operation. The Li-CO$_2$ cell under investigation consists of several components: a lithium metal sheet serving as the negative electrode, a separator, a porous carbon cathode coated with Sb$_{0.67}$Bi$_{1.33}$Te$_3$, and a liquid electrolyte that permeates the porous structures, as illustrated schematically in Figure \ref{Figure2}. The used electrolyte composition is described in section \ref{MD}. To facilitate electron transfer, current collectors are located at the rear of both electrodes. This configuration enables a thorough examination of the electrochemical and transport processes that govern battery cell performance and limitations.
The main objective of this mathematical framework is to provide a comprehensive understanding of the intricate interactions among various components and processes within the Li-CO$_2$ battery cell system.
\subsubsection{Electrode Reactions}
\subsubsection*{Cathode Reaction}
The porous carbon electrode provides a site for the electrochemical reduction of carbon dioxide. During operation, CO$_2$ from the external feed side dissolves in the electrolyte, moves through the pores of the positive electrode, and reacts with lithium ions at the active sites. The reaction considered at the positive electrode is \cite{mu2024toward}:
\begin{equation} \label{eq:4}
3\mathrm{CO}_2 + 4\mathrm{Li}^+ + 4\mathrm{e}^- \rightleftharpoons 2\mathrm{Li}_2\mathrm{CO}_3 + \mathrm{C} \qquad (\mathrm{E}^0_d=2.8\mathrm{V},\qquad \mathrm{E}^0_c=3.8\mathrm{V})
\end{equation}
where $\mathrm{E}^0_d=2.8 V$ and $\mathrm{E}^0_c=3.8 V$ represent the equilibrium potentials for discharge and charge reactions, respectively \cite{wang2023developing, zhang2022challenges}. The reaction products (Li$_2$CO$_3$ and C) have limited solubility in the DMSO-based liquid electrolyte (0 mol/L for C and 0.0371 mol/L for Li$_2$CO$_3$). When their concentrations exceed these solubility limits, the electrolyte solution becomes supersaturated \cite{welland2015atomistically}, leading to the deposition of Li$_2$CO$_3$ and C. This is followed by the nucleation and growth of these products in the liquid phase within the porous electrode. These deposited products significantly impact the battery's performance by covering the active surface area, obstructing pathways for reactive species (Li$^+$ and CO$_2$), and inhibiting further reactions within the cathode, thereby limiting the extent of cell discharge.
\subsubsection*{Anode Reaction}
The main reaction at the anode, which represents the reversible deposition and dissolution of lithium metal during battery cell operation, is described as \cite{FEA2}:
\begin{equation}
\mathrm{Li}^+ + \mathrm{e}^- \rightleftharpoons \mathrm{Li} \qquad (\mathrm{E}^0=0.0\mathrm{V})
\end{equation}
Where $\mathrm{E}^0$ is the reaction equilibrium potential for charge and discharge \cite{FEA2}.\\
\subsubsection{Particle transport at the macroscopic level}
The Nernst-Planck equation,  a continuity equation for the time-dependent concentration $c_i(t,{\bf x})$ of a chemical species, is used to describe the transport of chemical species within the battery cell:
\begin{equation}
{{\partial \varepsilon c_i} \over {\partial t}} + \nabla \cdot {\bf J_i} = r_i
\end{equation}
In this context, \(\varepsilon\) represents the porosity of the electrode or the electrolyte region in the solution phase, \({\bf J_i}\) denotes the flux, and \(r_i\) indicates the volumetric production rate of species \(i\) transitioning from the electrode material to the electrolyte within the porous structure. The total flux is composed of three components: diffusion, advection, and electromigration. This implies that the concentration is affected by an ionic concentration gradient $\nabla c$, flow velocity {\bf v}, and an electric field {\bf E}.
\begin{eqnarray}
    {\bf J} = -\underbrace{D^{eff}\nabla c}_{\text{Diffusion}} + \underbrace{c{\bf v}}_{\text{Advection}}+\underbrace{{D^{eff}ze\over{k_\text{B}T}}c{\bf E}}_{\text{Electromigration}}
\end{eqnarray}

In this equation, $D^{eff}$ denotes the effective diffusion coefficient of the chemical species, $z$ represents the valence of the ionic species, $e$ represents the elementary charge, $k_\text{B}$ represents the Boltzmann constant, and $T$ represents the absolute temperature. The electric field may be further decomposed as:
\begin{equation}
    {\bf E} = -\nabla \phi - {\partial {\bf A}\over{\partial t}}
\end{equation}
where $\phi$ is the electric potential and 
$A$ is the magnetic vector potential. Therefore, the Nernst–Planck equation is given by:
\begin{equation}
     \frac{\partial c}{\partial t} = \nabla \cdot \left[ D^{eff}\nabla c - c\mathbf{v} + \frac{D^{eff}ze}{k_\text{B}T} c \left( \nabla \phi + {\partial {\bf A}\over{\partial t}} \right) \right] 
\end{equation}

The primary transport mechanisms in batteries, ionic migration due to electric fields and diffusion due to concentration gradients, are accurately described by the Nernst-Planck equation, which excludes time-varying magnetic effects \cite{costa2021magnetically}:

\begin{equation}
     \frac{\partial c}{\partial t} = \nabla \cdot \left[ D^{eff}\nabla c - c\mathbf{v} + \frac{D^{eff}ze}{k_\text{B}T} c \nabla \phi \right] 
\end{equation}

The moving boundaries theory of transference number ($t_+$) gives \cite{macinnes1932transference}:
\begin{equation}
    t_+ = \frac{vc_+F}{I\Delta t}
\end{equation}
where $v$ is the volume occupied by diffusing cations by the boundary in time $\Delta t$, $c_+$ the Li$^{+}$ concentration, F the Faraday constant, and $I$ the electric current.\\
By introducing the electrode surface $S$:
\begin{equation}
    t_+ = \frac{vc_+F/S}{I\Delta t/S}
\end{equation}
Also, we can write $c$ as:
\begin{equation}
    c=\frac{N}{N_A v}
\end{equation}
with $N$ the number of particles and $N_A$ the Avogadro number.\\
Then:
\begin{equation}
    t_+ = \frac{\textbf{J}F}{{\bf i_{2}}}
\end{equation}

with,
$\textbf{J}$ the molar flux of cations, $i_2=I/S$ the current density in the electrolyte, and $F$ the Faraday constant.

\begin{table}[ht!]
\caption{\normalsize Parameters used in the Finite Element Analysis (FEA).}

  \label{Table2}
\begin{tabular*}{\textwidth}{@{\extracolsep{\fill}}llll}
    \hline

Symbol  &  Value &  Description &  Reference\\
     \hline 
 \multicolumn{4}{l}{Parameters used in the battery assembly} \\
L$_{sep}$ & 50 $\mu$ m & Thickness of Separator & This work \\
L$_{pos}$ & 2 mm & Thickness of positive electrode & This work\\
A$_c$ & 1 cm$^2$ & Cross-sectional area & This work\\
$\varepsilon_{l,0}$ & 0.73 & Initial porosity of positive electrode & This work\\ 
m$_{Sb_{0.67}Bi_{1.33}Te_3}$ & 0.4 mg & mass of active cathode material & This work\\
$\varepsilon_{sep}$ & 0.5 & Porosity of separator &\\
$\varepsilon_{s,0}$ & 1-$\varepsilon_{l,0}$ & Initial active material solid fraction of cathode & This work\\
c$_{Li,0}$ & 1000 mol/m$^3$ & Initial concentration of Li$^+$ in electrolyte & This work\\
\multicolumn{4}{l}{Parameters calculated using DFT/AIMD simulation} \\
$\kappa_{pos}$ & 1.9 10$^5$ S/m & Conductivity of positive electrode & This work\\
$\kappa_{Li_{2}CO_{3}}$ & 1.8 10$^-9$ S/m & electrical conductivity of Li$_{2}$CO$_{3}$ film & This work\\
\multicolumn{4}{l}{Parameters calculated using MD simulation} \\
D$_{Li^{+}}$ & 3.17 10$^{-9}$ m$^2$/s & Diffusion coefficient of Li$^+$ in electrolyte & This work\\
D$_{CO_2}$ & 1.2 10$^{-9}$ m$^2$/s & CO$_2$ diffusion coefficient & This work\\
$\sigma$ & 6.77 mS/cm & Conductivity of Li$^+$ in electrolyte & This work\\
t$_+ $ & 0.91 & Transference number of Li$^+$ in electrolyte & This work\\   
\multicolumn{4}{l}{General parameters} \\
i$_{0,Li}$ & 0.965 A/m$^2$ & Exchange current density for anode & \cite{li2012optimization, FEA2}\\
dlnf$_\pm$/dlnc & -1.03 & Activity dependence & \cite{li2012optimization, FEA2}\\
$\mathrm{E}^0_d$ & 2.8 V & Equilibrium potential for discharge & \cite{wang2023developing}\\
$\mathrm{E}^0_c$ & 3.8 V & Equilibrium potential for charge & \cite{zhang2022challenges}\\
r$_{pos,0}$ & 25 nm & Particle radius in the positive electrode & \cite{FEA2}\\
a$_{0}$ & 3$\varepsilon_{s,0}$/r$_{pos,0}$ & Initial active specific surface area & \cite{FEA2}\\
n & 4 & Number of transferred electrons  & \cite{wang2023developing}\\
c$_{Li_{2}CO_{3},max}$ & 0.0371 mol/L & Solubility limit of Li$_{2}$CO$_{3}$ in DMSO & \\
c$_{C,max}$ & 0 mol/L & Solubility limit of Carbon in DMSO & \\
$\rho_{Li_{2}CO_{3}}$ & 2140 kg/m$^3$ & Density of Li$_{2}$CO$_{3}$ & \\
M$_{Li_{2}CO_{3}}$  & 73.89 10$^{-3}$ kg/mol & Molecular weight of Li$_{2}$CO$_{3}$ & \\
$\rho_C$ & 2.11 g/cm$^3$ & Density of carbon  & \\
T & 300 K & Temperature & \\
\multicolumn{4}{l}{Assumed parameters} \\
k$_a$ & 1.11 10$^{-15}$ m/s & Anodic reaction rate coefficient& Assumed\\
k$_c$ & 3.14 10$^{-15}$ m$^7$/s/mol$^2$ & Cathodic reaction rate coefficient& Assumed\\
c$_{CO_{2,0}}$ & 4.8 mol/m$^3$ & Initial CO$_2$ concentration in positive electrode & Assumed\\ 
 \hline
\end{tabular*}
\label{Table1}
\end{table}

\begin{table*}[ht!]
\caption{\normalsize Boundary conditions at boundaries 1 (x=0), 2 (x=L$_{sep}$), and 3 (x=L$_{sep}$ + L$_{pos}$).}
\begin{tabular*}{\textwidth}{@{\extracolsep{\fill}}lll}
\hline
 Variable  & Value & Boundary \\
     \hline 
Total current density & i$_{app}$ & 3 \\
c$_{CO_2} $ & c$_{CO_{2,0}} $ & 3 \\
$\phi_{1}$ & 0 V & 1  \\
$\nabla c_{CO_2}$ & 0 & 2 \\
$\nabla c_{Li}$ & 0 & 1, 3 \\
 \hline
\end{tabular*}
\label{Table2}
\end{table*}

\begin{table*}[ht!]
\caption{\normalsize Initial values of independent and dependent variables.}
\begin{tabular*}{\textwidth}{@{\extracolsep{\fill}}lll}
\hline
Variable  & Value & Domain \\
     \hline 
$\phi_{1}$ & 2.8 V & Porous cathode  \\
$\phi_{1}$ & 0 V & Separator  \\
$\phi_{2}$ & 0 V & Porous cathode, Separator  \\
$c_{Li}$ & c$_{Li,0}$ & Porous cathode, Separator \\
$c_{CO_2}$ & c$_{CO_{2,0}}$ & Porous cathode \\    
$c_{Li_{2}CO_{3}}$ & c$_{Li_{2}CO_{3},max}$ & Porous cathode\\
$c_{C}$ & c$_{C,max}$ & Porous cathode \\
$\varepsilon$ & $\varepsilon_{l,0}$ & Porous cathode \\
 \hline
\end{tabular*}
\label{Table3}
\end{table*}

The primary interest in battery modeling is electrochemical processes rather than fluid dynamics, so the advection term is considered less critical \cite{meng2014numerical}. Furthermore, the porous structure limits large-scale fluid motion, further reducing the importance of advection. By neglecting the Advection term, the Nernst–Planck equation for Li$^+$ and CO$_2$ can be expressed as:
\begin{eqnarray}
     {\bf J_{Li^+}} = -D^{eff}_{Li^+}\nabla c_{Li^+} + \frac{{\bf i_{2}}t_+}{F} \\
     {\bf J_{CO_2}} = -D^{eff}_{CO_2}\nabla c_{CO_2}
\end{eqnarray}
The effective diffusion coefficient of a species i (i=Li$^+$, CO$_2$) was calculated using the Bruggeman equation \cite{mehta2021li, mukouyama2023finite}:
\begin{eqnarray}
     D^{eff}_{i}=\varepsilon^{1.5} D_{i}     
\end{eqnarray}
The $\varepsilon^{1.5}$ form, in the Bruggeman relation, emerges from the mathematical treatment of a system of spherical particles dispersed in a continuous medium \cite{usseglio2018resolving}. This form has been found to provide a reasonable empirical fit approximation for many real porous systems, especially at higher porosities \cite{usseglio2018resolving, holzer2023tortuosity}. The diffusion coefficient $D_{i}$ (i=Li$^+$, CO$_2$) is calculated from our MD simulation.
\subsubsection{Equations of battery electrochemical properties}
The surface density of the liquid phase ${\bf i_{2}}$ represents the function of the electrolyte concentration $c_{Li}$, time $t$, the ionic conductivity $\sigma^{eff}$, the absolute temperature $T$, the activity coefficient $\frac{{\partial \ln f_{ \pm } }}{{\partial \ln c_{Li} }}$, the universal gas constant $R$, the cathode porosity $\varepsilon$ and electrode solution-phase potential $\phi_{2}$ \cite{FEA1, FEA2}. 
\begin{equation}
    {\bf i_{2}} = - \sigma^{eff} \nabla \phi_{2} + \left( {\frac{2\sigma^{eff}RT}{F} + \frac{2\sigma^{eff}RT}{F}\frac{{\partial \ln f_{ \pm } }}{{\partial \ln c_{Li} }}} \right)\left( {1 - t_{+} } \right)\nabla \ln c_{Li}
\end{equation}
\begin{equation}
\sigma^{eff}=\varepsilon^{1.5} \sigma
\end{equation}
The solid phase current density ${\bf i_{1}}$ is the function of the solid-electrode potential $\phi_1$ and the electronic conductivity of solid-phase mixture $\kappa_{pos}$ \cite{FEA1, FEA2},
\begin{equation}
   {\bf i_{1}} = - \kappa_{pos}^{eff} \nabla \phi_{1} 
\end{equation}
\begin{equation}
\kappa_{pos}^{eff}=(1-\varepsilon_{l,0})^{1.5} \kappa_{pos}
\end{equation}
$\varepsilon_{l,0}$ denotes the starting porosity of the cathode. It is important to mention that the effect of deposits (Li$_2$CO$_3$ and C) with low conductivity on $\kappa_{pos}$ was not taken into account, and it was assumed that $\kappa_{pos}$ would not change even as the battery reactions occurred. This assumption is primarily justified by the demonstrated structural and chemical stability of the catalyst under cycling conditions \cite{jaradat2024fast}. In our continuum model, we further decouple electron transport through the deposited Li$_2$CO$_3$ and carbon byproducts, treating their intrinsic conductivities separately from that of the catalyst. This approach ensures that the catalyst’s role in maintaining electronic percolation pathways is accurately represented. However, it is important to acknowledge the potential limitations of this simplification. If the catalyst were not stable-for example, if it were susceptible to pulverization or surface passivation-its conductivity could decrease during cycling. Such changes might result in localized current hotspots or capacity fade, which would not be captured by the current model. Therefore, while the assumption is reasonable for stable catalyst systems, it may not be universally applicable to all materials or operational conditions.

\subsubsection*{Conservation of charge }
\begin{equation}
     \nabla {\bf i_{1}}+\nabla {\bf i_{2}}=0 
\end{equation}
\begin{equation}
   \sum_{i} s_{i} M_{i}^{z_{i}} \rightarrow n e^{-}
\end{equation}
\begin{equation}
     \nabla \cdot {\bf i_{2}}=aj
\end{equation}

These equations demonstrate that the transfer current per unit electrode volume corresponds to the electrode chemical reaction rate, where $M_{i}$ is a symbol representing a species participating in the electrochemical reaction, $z_{i}$ and $s_{i}$ are the charge number and the stoichiometric coefficient of species $i$, respectively, $n$ is the number of electrons transferred in the reaction, $a$ is the specific interfacial area of the pore per unit electrode volume, and $j$ is the average transfer current. In the present work, only one electrochemical reaction is considered: the formation of $\mathrm{Li}_2\mathrm{CO}_3$/carbon ($\mathrm{Li}_2\mathrm{CO}_3$/C) inside the porous cathode. In practical cells, battery side reactions can occur during cycling alongside the main electrode reactions. This is reflected by the superficial production rate $r_{i}$ of a species from the cathode's solid phase to the pore solution, given by Faraday's law as a function of the local transfer current density $j_c$ between the cathode/electrolyte interface.
\begin{equation}
     r_{i}=-\frac{as_i}{nF}j_c
\end{equation}

\subsubsection*{Rate expressions}
For modeling the cathode's electrochemical reaction, a modified version of the Butler-Volmer equation is applied in the model using two rate coefficients. This modification is necessary because the cathode reaction presented in equation \ref{eq:4} depends on both the concentrations of Li$^+$ and CO$_2$, as well as the concentration of Li$_2$CO$_3$/C \cite{zhang2024modeling}:
\begin{equation}
     \frac{j_c}{nF} =  k_a (c_{\mathrm{Li}_2\mathrm{CO}_3/C,s}) \exp \left[ \frac { (1-\beta) nF } {RT} \eta_{\rm c} \right] - k_c (c_{\mathrm{Li}^+,s})^2 (c_{CO_2,s})\exp \left[ \frac { -\beta nF } {RT} \eta_{\rm c} \right] 
\end{equation}

\begin{equation}
    \eta_{\rm c} = \phi_{1} - \phi_{2}-\Delta \phi_{film}-E^0
\end{equation}
\begin{equation}
    \phi_{film} = j_c R_{film} \varepsilon_s
\end{equation}
where $j_c$ is local transfer current density between electrode and electrolyte interface, $c_{i,s}$ is the molar concentration of species $i$ at the cathode surface, $k_a$ and $k_c$ are the anodic and cathodic rate constant, respectively, $\beta$ is a symmetry factor equal to 0.5, $\eta_{\rm c}$ is surface or activated overpotential for reaction at cathode, $\phi_{film}$ and $R_{film}$ are the voltage drop and the electrical resistivity across $\mathrm{Li}_2\mathrm{CO}_3$/C film formation, respectively, $\varepsilon_s$ is the volume fraction of solid $\mathrm{Li}_2\mathrm{CO}_3$/C, and $E^0$ is the theoretical open-circuit potential for reaction.\\

The properties of the $\mathrm{Li}_2\mathrm{CO}_3$/C film, specifically its electrical resistivity $R_{film}$, thickness ($\Delta s$), and conductivity ($\kappa_{film}$), are related by the equations below:
\begin{equation}
    R_{film} = \frac{\Delta s}{\kappa_{film}}
\end{equation}
\begin{equation}
\Delta s=\Delta s_C + \Delta s_{\mathrm{Li}_2\mathrm{CO}_3}
\end{equation}
\begin{equation} \label{eq:31}
\Delta s_i=\frac{\Delta \varepsilon_i}a=\frac{M_i c_i}{a\rho_i} 
\end{equation}
where $i$ represents $\mathrm{Li}_2\mathrm{CO}_3$ or C, while $\Delta s_i$, $\Delta \varepsilon_i$, $M_i$, $a$, and $\rho_i$ refer to the film thickness of $i$, the change in porosity of the cathode caused by $i$, the molar mass of $i$, the specific area, and the density of $i$, respectively.\\

The electrochemical reaction at the lithium metal anode is given by the general Butler-Volmer equation as follows \cite{FEA1, FEA2}:
\begin{equation}
     j_a = i_0 \exp \left[ \frac { (1-\beta) nF } {RT} \eta_{\rm a} \right] -\exp \left[ \frac { -\beta nF } {RT} \eta_{\rm a} \right] 
\end{equation}
where $i_0$ is the exchange current density for the anode, $\eta_{\rm a}$ is the surface or activated overpotential for the reaction at the anode, and the other parameters are as described above.

The specific area $a$ of the electrode/electrolyte interface in Equation \ref{eq:31} is decreased by the morphology and dynamic change in porosity due to $\mathrm{Li}_2\mathrm{CO}_3$/C solid passivation during the electrochemical reaction. This effective local surface area per unit volume of the electrode can be commonly written by a geometric relation \cite{FEA1, FEA2}:
\begin{equation}
     a = a_0 \exp \left[ 1-\frac{\varepsilon_{\mathrm{Li}_2\mathrm{CO}_3/C}}{\varepsilon_{l,0}} \right] 
\end{equation}
where $a_0$ the initial specific surface area, $\varepsilon_{\mathrm{Li}_2\mathrm{CO}_3/C}$ and $\varepsilon_{l,0}$ are the volume fraction of solid $\mathrm{Li}_2\mathrm{CO}_3$/C and initial electrode porosity, respectively.

\subsubsection*{Transport through $\mathrm{Li}_2\mathrm{CO}_3$/C layer}
During discharge, the formed $\mathrm{Li}_2\mathrm{CO}_3$/C creates an additional barrier for Li$^+$ and $\mathrm{CO}_2$ diffusion from the porous medium to the cathode surface.

This process follows Fick's law \cite{FEA2}, where the flux is proportional to the concentration gradient between the bulk electrolyte and the active surface, sustained by ongoing electrochemical reactions consuming reactants.
\begin{equation}
     -\frac{as_i}{nF}j_c=\frac{aD^{eff}_{i,film}(c_i-c{i,s})}{l} 
\end{equation}
where $D^{eff}_{i,film}$ is the effective diffusion coefficient of species $i$ across the $\mathrm{Li}_2\mathrm{CO}_3$/C film, and $l$ is the thickness of the film. $c_{i}$ represents the molar concentration of species $i$ in the electrolyte, while $c_{i,s}$ denotes the molar concentration of species $i$ specifically at the cathode surface. The effective diffusivity for both $\mathrm{Li}^+$ and CO$_2$ reactants is used to describe how the pores are deposited by the $\mathrm{Li}_2\mathrm{CO}_3$/C formation; this can be described by the Bruggeman relation \cite{mehta2021li, mukouyama2023finite}:
\begin{equation}
     -\frac{\partial \varepsilon}{\partial t}=\frac{a j_{c} M_{\mathrm{Li}_2\mathrm{CO}_3/C}}{2F\rho_{\mathrm{Li}_2\mathrm{CO}_3/C} }
\end{equation}
The volume fraction of the solid $\mathrm{Li}_2\mathrm{CO}_3$/C and the thickness of the film ($l$) can be determined from the cathode volume balance as:
\begin{equation}
    \varepsilon_{\mathrm{Li}_2\mathrm{CO}_3/C}=1-\varepsilon -\varepsilon_{l,0}
\end{equation}

\begin{equation}
 l= \left[ \frac{\varepsilon_{\mathrm{Li}_2\mathrm{CO}_3/C}+\varepsilon_{s,0}}{\varepsilon_{s,0}} \right]^{(1/3) }  r_0-r_0
\end{equation}

In this context, $M_{\mathrm{Li}_2\mathrm{CO}_3/C}$ and $\rho_{\mathrm{Li}_2\mathrm{CO}_3/C}$ represent the molar mass and density of ${\mathrm{Li}_2\mathrm{CO}_3/C}$, respectively. The term $\varepsilon_{s,0}$ denotes the volume fraction of the initial solid phase of the cathode material, while $r_0$ indicates the radius of the particles within the electrode.
\subsubsection*{Initial conditions/Boundary conditions}
The governing equations consist of partial differential equations that include time and space as independent variables. The dependent variables were simulated using COMSOL Multiphysics 6 Finite Element Analysis software to examine their spatial and temporal changes. The battery cell was segmented into three regions: the anode, separator, and porous cathode. An extra fine mesh was selected for enhanced accuracy. To reduce computation time, a variable-step solver was employed with a relative tolerance of 1 $\times$ 10$^{-4}$. This solver decreases the total number of time steps required for the calculation. The parameters used in the simulations are presented in Tables \ref{Table1},  \ref{Table2},  and \ref{Table3}. The applied current density, i$_{app}$, was considered positive when directed from the cathode to the external circuit (Figure \ref{Figure2}). The total current density passing through boundary 3 was designated as i$_{app}$. The Li$^+$ flux at boundaries 1 and 3 (J$_{Li}$) and CO$_2$ flux at boundary 2 (J$_{CO_2} $) were set to zero. Additionally, the boundary condition $\nabla$c$_i$ = 0 was applied to the relevant boundaries (see Table \ref{Table2}). The parameter c$_{CO_2} $ was assumed to be equal to the saturation concentration, c$_{CO_2,0}$, at boundary 3. The boundary conditions are outlined in Table \ref{Table2}. Discharge simulations were conducted with applied current densities (i$_{app}$) ranging from 0.1 to 1 mA/cm$^{2}$, defined as positive in the direction from the cathode to the external circuit.
\section{Results and discussion} \label{Results}
\subsection{Electrical conductivity from DFT, AIMD, and Kubo-Greenwood formula} 
\begin{figure*}[!h]
	\begin{center}
		\begin{minipage}[t]{1\textwidth}
					\centering
			\includegraphics[width=1\textwidth]{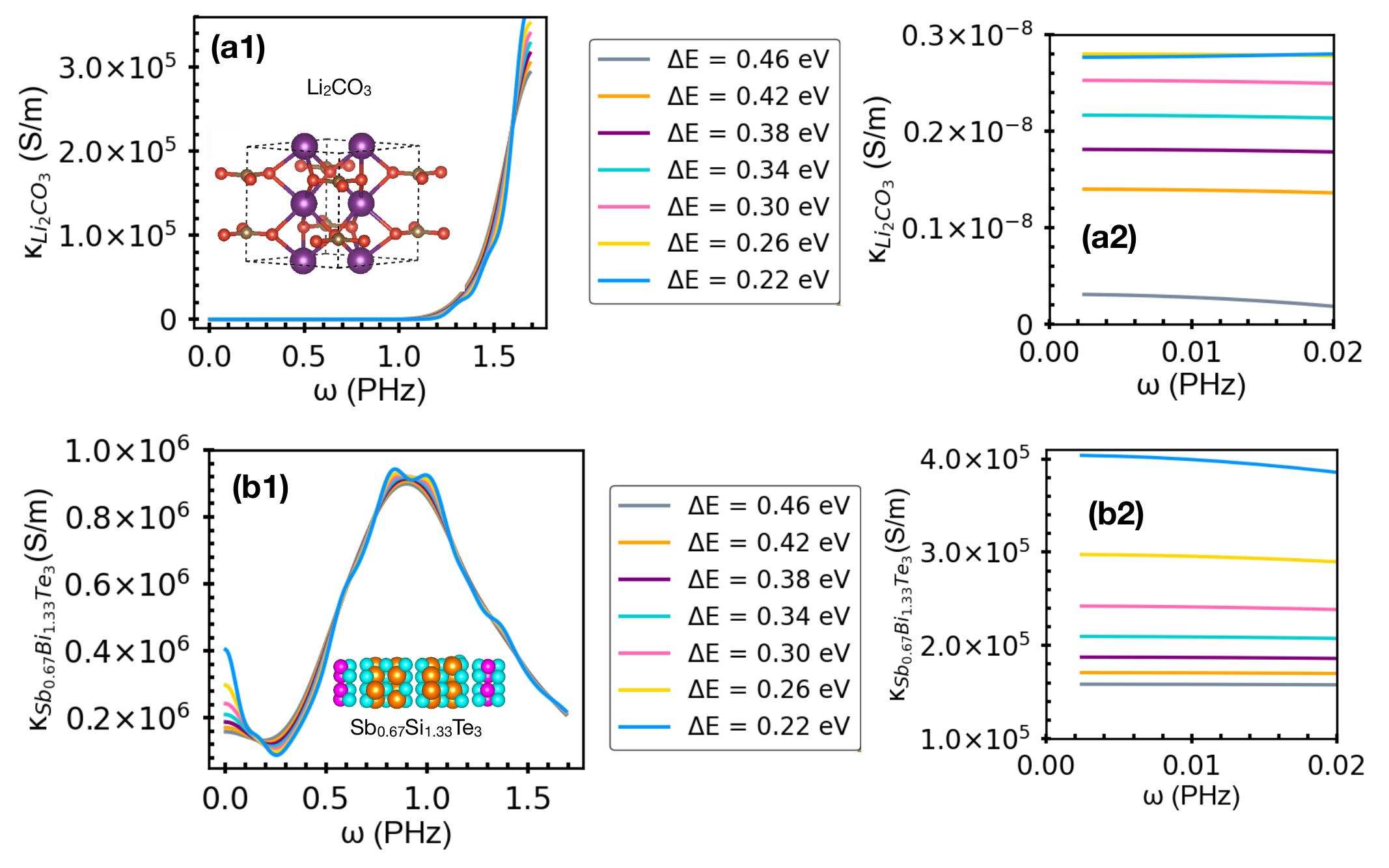}
            \caption{ Dynamic electrical conductivity of Li$_2$CO$_3$ (a1-a2) and Sb$_{0.67}$Bi$_{1.33}$Te$_3$ (b1-b2) calculated using the Kubo-Greenwood formula for a wide frequency range (a1 and b1) and a small frequency range (a2 and b2). The electrical conductivity is plotted using $\Delta$E values from 0.22 to 0.46 eV.} 
            \label{Figure3} 
		\end{minipage}
	\end{center}
\end{figure*}

The calculated dynamic electrical conductivities for Sb$_{0.67}$Bi$_{1.33}$Te$_3$ and Li$_2$CO$_3$ are shown in Figure \ref{Figure3} (a1) and (b1) across a broad frequency range, and in Figures \ref{Figure3} (a2) and (b2), which focus on low frequencies to enable extrapolation to 0 Hz, with $\Delta$E values ranging from 0.22 eV to 0.46 eV. Static electrical conductivities at zero frequency were determined by extrapolating the low-frequency data in Figures \ref{Figure3} (a2) and (b2). The calculated conductivities are within the experimental order of magnitude. $\Delta$E values between 0.34 eV and 0.42 eV were selected because they show convergence with experimental conductivity data within this range. Furthermore, varying the electrical conductivity within the computational uncertainty does not significantly affect the predicted battery properties in the FEA. The static electrical conductivities calculated for different values of $\Delta$E show the conductive behavior of Sb$_{0.67}$Bi$_{1.33}$Te$_3$ (a conductivity of the order of $10^5$ S/m) and insulative behavior of Li$_2$CO$_3$ (a conductivity of the order of $10^{-9}$ S/m). Our computational analysis reveals a static electrical conductivity of $(1.9 \pm 0.3) \times 10^5$ S/m for Sb$_{0.67}$Bi$_{1.33}$Te$_3$, indicating its strong electrical conducting properties. This finding aligns with the well-established characteristics of bismuth telluride (Bi$_2$Te$_3$) and its alloys, which are extensively employed in thermoelectric devices due to their good electrical conductivity \cite{goldsmid2014bismuth, park2016thermal}. The high conductivity can be attributed to the semiconductor nature of Bi$_2$Te$_3$ and the antimony (Sb) doping effect, which augments the carrier concentration \cite{lou2022tunable}. Experimental investigations on Bi$_2$Te$_3$ have reported electrical conductivity values as high as $1.71 \times 10^5$ S/m at 323 K \cite{lou2022tunable}, further corroborating the computational results. Our calculations reveal that Li$_2$CO$_3$ exhibits a static electrical conductivity of $(1.8 \pm 0.3) \times 10^{-9}$ S/m at 300 K, confirming its highly insulating nature. This result closely aligns with experimental measurements of $8 \times 10^{-9}$ S/m at 393 K \cite{ghobarkar1993hydrothermal}. The slight difference between calculated and experimental values is consistent with the expected increase in conductivity at higher temperatures, underscoring the reliability of our computational approach. The insulating nature of Li$_2$CO$_3$ can be attributed to its ionic nature and the strong ionic bonds that hinder electron mobility \cite{garcia2013dft+}. In Li$_2$CO$_3$, Li$^+$ ions act as the primary charge carriers rather than electrons, a characteristic typical of ionic compounds. The low conductivity of Li$_2$CO$_3$ presents challenges to battery performance, including high internal resistance and passivation of the electrode surface \cite{mao2022carbon, tang2021promoting}.\\
The calculated electrical conductivities are used in the continuum model to characterize electron percolation in the cathode and the deposited Li$_2$CO$_3$ during the battery charge/discharge cycle.

\subsection{Surface reaction of Sb$_{0.67}$Bi$_{1.33}$Te$_3$ as cathode catalyst}
\begin{figure*}[!ht]
	\begin{center}
		\begin{minipage}[t]{1\textwidth}
				\centering
			\includegraphics[width=0.9\textwidth]{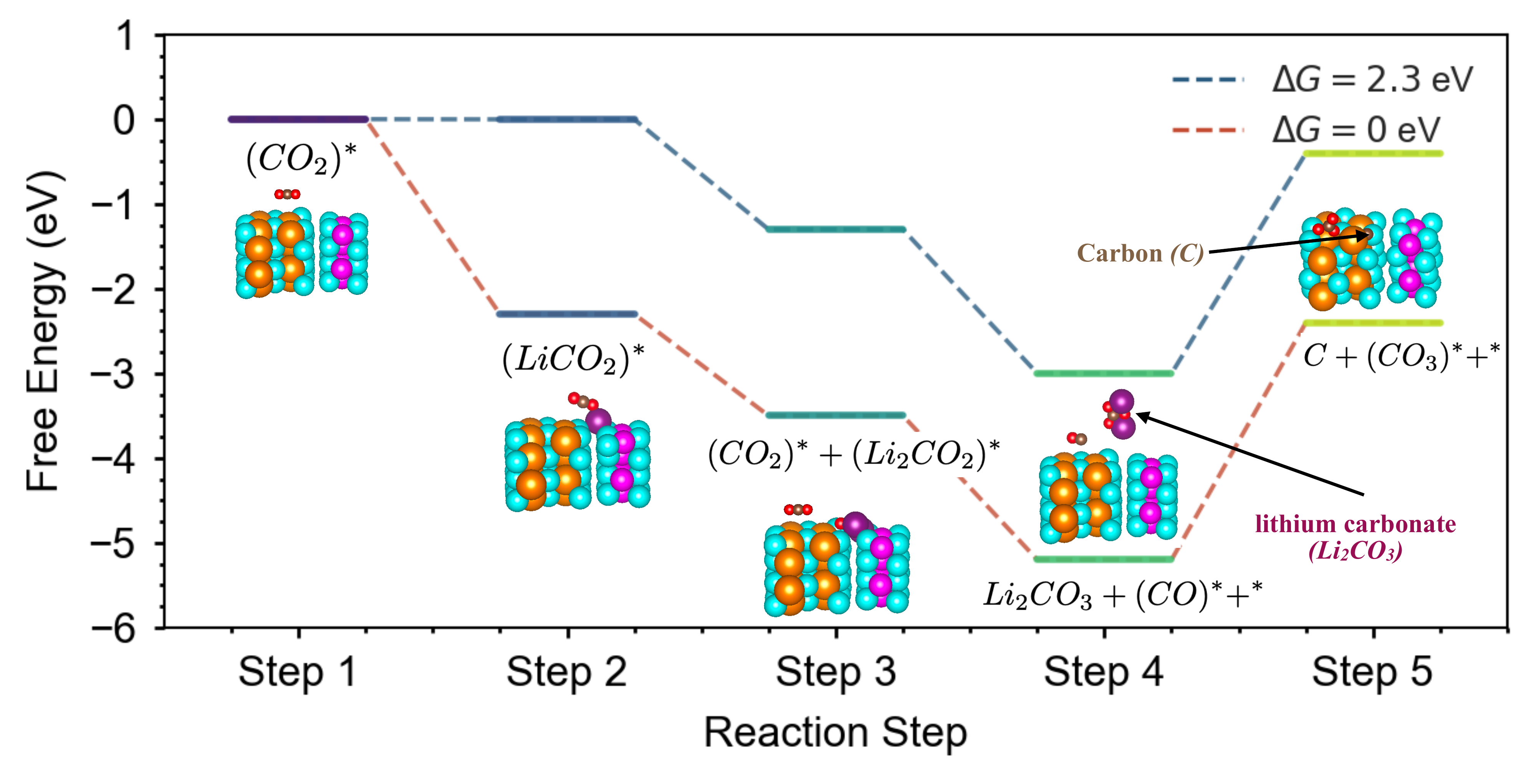}
			\caption{Free energy landscape of the CO$_2$ reduction reaction leading to the formation of Li$_2$CO$_3$/C deposits on the Sb$_{0.67}$Bi$_{1.33}$Te$_3$ surface, with inset images illustrating the optimized geometries of key reaction intermediates.}
           \label{Figure4} 
		\end{minipage}
	\end{center}
\end{figure*}

To comprehensively understand the formation of Li$_2$CO$_3$ and C on the Sb$_{0.67}$Bi$_{1.33}$Te$_3$ catalyst surface, we calculated the Gibbs free energy ($\Delta G$) changes for various elementary reaction steps on this surface as shown below (steps 1-5) similar to the calculations done in Ref. \cite{jaradat2024fast}. Here, we used a slightly bigger supercell (as shown in Figure \ref{Figure3} (a1)) compared to the work done in Ref. \cite{jaradat2024fast}. The reaction pathway revealed a sequence of intermediate stages, beginning with the adsorption of CO$_2{}^*$ on a Bi site with a Bi-O distance of 3.5 \AA, which was marginally endergonic ($\Delta G = 0.002$ eV). The coupled addition of Li$^+$ and an electron to the adsorbed CO$_2{}^*$ formed the LiCO$_2{}^*$ intermediate, a highly exergonic step ($\Delta G = -2.3$ eV). Subsequent interactions between two LiCO$_2{}^*$ intermediates produced Li$_2$CO$_2{}^*$ and CO$_2{}^*$, with a $\Delta G$ of $-1.2$ eV.
Further, the conversion of Li$_2$CO$_2{}^*$ and CO$_2{}^*$ into Li$_2$CO$_3$ and CO${}^*$ ($\Delta G = -1.7$ eV) marked a pivotal step in the pathway. Finally, the generation of carbon was facilitated through the reaction of CO and CO$_2$, an energetically uphill step ($\Delta G = 2.8$ eV) that also produced CO$_3{}^*$ on the surface. These results, illustrated in Figure \ref{Figure4}, highlight the essential role of coupled electron-cation transfer processes in the growth of discharge products. The results in this paper are consistent with the work reported in Ref. \cite{jaradat2024fast}.
The Gibbs free energy \( G \) of the reaction was calculated using the following equation:
\begin{equation}
G = E + \text{ZPE} + \int_0^T C_p \, dT - TS
\end{equation}
where ZPE is the zero-point energy, $C_p$ is the heat capacity, $S$ is the entropy, $T$ is the temperature, and $E$ is the electronic energy.\\

In the calculation of Gibbs free energy profiles, full corrections including zero-point energy (ZPE) and entropy were applied explicitly to the initial gas-to-surface adsorption step of CO$_2$, where such contributions are known to be significant due to the loss of translational and rotational entropy. For subsequent steps involving transformations between adsorbed intermediates (e.g., Li$_2$CO$_3$, Li$_2$CO$_2$, Li$_2$CO$_3$, and C–CO$_3$ species), we assumed that ZPE and entropy effects largely cancel out, as these involve similar bonding environments and restricted surface-bound species. This approach is widely used in heterogeneous catalysis modeling and has been shown to introduce minimal quantitative error, while preserving the qualitative accuracy of energy landscapes and mechanistic conclusions \cite{norskov2004origin}. Accordingly, our analysis emphasizes the dominant electronic energy contributions for these surface processes, which govern the reaction thermodynamics and barrier-lowering effects central to catalyst performance. The potential-dependent reaction energies were calculated based on the computational hydrogen electrode (CHE) model with reference to the Li/Li$^+$ electrode \cite{abild2007scaling, peterson2010copper}. The potential-dependent reaction energies profile is shown in Figure \ref{Figure4}, and the respective reaction energies of each step are shown in Table \ref{Table4}.
To level up the electrochemical step, a potential was applied to the formation of LiCO$_2$. As can be seen in Figure \ref{Figure4}, this potential occurs at 2.3 V versus Li/Li$^+$ electrode \cite{jaradat2024fast}.\\
The complete 5-step reaction pathway can be summarized as follows:
\begin{align*}
&^*+CO_2 (g) \rightarrow CO_2{}^* &\qquad& \text{(Step-1)} \\
&O_2{}^*+Li^++e^- \rightarrow LiCO_2{}^* &\qquad& \text{(Step-2)} \\
&2LiCO_2{}^*\rightarrow CO_2{}^*+Li_2CO_2{}^* &\qquad& \text{(Step-3)} \\
&CO_2{}^*+Li_2 CO_2{}^*\rightarrow Li_2 CO_3+CO{}^*{}+{}^* &\qquad& \text{(Step-4)} \\
&CO_2{}^*+CO{}^*\rightarrow C+CO_3{}^*{}+{}^* &\qquad& \text{(Step-5)}
\end{align*}

The strong exergonic nature of steps 2-4 suggests that these processes occur readily, while the endergonic nature of the final carbon formation step aligns with the observed preferential adsorption of carbon on the Sb$_{0.67}$Bi$_{1.33}$Te$_3$ catalyst surface. The Li$_2$CO$_3$ is formed in step 4, but this step also releases CO${}^*$ and a free surface site (${}^*$). This suggests that Li$_2$CO$_3$ formation occurs near, but not directly on, the catalyst surface, reducing the cathode porosity and possibly clogging the cathode pores.\\
These insights into the reaction mechanism have significant implications for catalyst design and overall battery performance. The strong affinity of carbon for the catalyst surface, coupled with the energetically unfavorable step of formation, suggests that catalyst degradation through carbon accumulation could be a concern over long-term battery operation. This highlights the need for strategies to mitigate carbon build-up, such as developing self-cleaning catalysts or implementing periodic regeneration protocols.
Furthermore, the spatial separation of Li$_2$CO$_3$ formation from the immediate catalyst surface underscores the importance of electrolyte engineering in Li-CO$_2$ battery design. Since Li$_2$CO$_3$ formation appears to occur away from the catalyst, the electrolyte's ability to facilitate Li$_2$CO$_3$ formation, dissolution, and transport becomes crucial.\\

The continuum model considers the formation of Li$_2$CO$_3$ and carbon in the cathode. We assume that the catalyst increases the reaction rate of Li$_2$CO$_3$/C formation and use a reaction rate constant on the order of 10$^{-15}$ m$^7$ s$^{-1}$ mol$^{-2}$. This value is higher than the typical value assumed for Li$_2$O$_2$ formation in Li-O$_2$ batteries with non-catalyzed carbon cathodes, which is on the order of 10$^{-17}$ m$^7$ s$^{-1}$ mol$^{-2}$ \cite{FEA2}.
\begin{table}[h]
    \centering
    \caption{Reaction energies of each of the elementary steps.}
    \begin{tabular}{@{\extracolsep{\fill}}lp{8cm}c} 
        \hline
        Reaction Step & Reaction Steps & \( \Delta G \) (eV) \\ 
        \hline
        Step-1 & \( \Delta G = G(\text{CO}_2{}^*) - G(^*) - G(\text{CO}_2,_{gas}) \) & 0.002 \\ 
        Step-2 & \( \Delta G \approx \Delta E = E(\text{LiCO}_2{}^*) - E(\text{Li}^+ + e^-) - E(\text{CO}_2{}^*) \) & -2.3 \\ 
        Step-3 & \( \Delta G \approx \Delta E = E(\text{Li}_2\text{CO}_2{}^*) + E(\text{CO}_2{}^*) - 2E(\text{LiCO}_2{}^*) \) & -1.2 \\ 
        Step-4 & \( \Delta G \approx \Delta E = E(\text{Li}_2\text{CO}_3) + E(^*) + E(\text{CO}{}^*) - E(\text{Li}_2\text{CO}_2{}^*) - E(\text{CO}_2{}^*) \) & -1.7 \\ 
        Step-5 & \( \Delta G \approx \Delta E = E(\text{C})+E({\text{CO}}3{}^*) + E(^*) - E(\text{CO}{}^*) - E(\text{CO}_2{}^*) \) & 2.8 \\ 
        \hline
    \end{tabular}
    \label{Table4}
\end{table}

\subsection{Li-CO$_{2}$ battery electrolyte transport properties and Li$^+$ solvation structure from large scale MD simulation.}
This section presents the results of MD simulations that elucidate the atomic-level dynamic and solvation properties of the Li-CO$_{2}$ battery electrolyte depicted in Figure \ref{Figure5} (a). These simulations provide microscopic insights into the electrolyte's behavior, bridging the gap between the atomic and macroscopic transport properties in our multiscale approach.

\subsubsection{Li-CO$_{2}$ battery electrolyte transport properties}
The primary goal is to investigate the electrolyte's dynamic properties governing ion and CO$_{2}$ transport mechanisms, including ionic conductivity, cation transference number, and CO$_{2}$ diffusion coefficient, as presented in Table \ref{Table5}. These properties directly impact the Li-CO$_{2}$ battery's electrochemical performance.\\
The cations' transference number is assessed by \cite{wang2019poly}:
\begin{equation}
	t_{+}=	\frac{D_{\mathrm{Li}^+}}{D_{\mathrm{Li}^+}+D_{\mathrm{TFSI}^-}} 
\end{equation}\\
Where the self-diffusion coefficients are determined from the mean-square-displacement of the center of mass presented in Figure \ref{Figure5} (b), calculated from the MD simulations as follows \cite{mohammed2021}:
\begin{equation}
\Delta^{2} \mathbf {r}_{CM} \left(t \right)=\langle \Delta^{2} \mathbf {r} \left(t \right) \rangle = \frac {1} {N} \sum_{i = 1}^{N} \left [\mathbf {r}_{i} \left (t + t_{0} \right) - \mathbf {r}_{i} \left (t_ {0} \right) \right]^{2}  \end{equation}
In the expression above, $ \mathbf{r}_{i} \left (t \right) $ represents the temporal position of a random walker (ion $i$ or CO$_{2}$ molecule). Here, $ t $ indicates the time, and $ t_{0} $ is the initial time at which the random walker begins to move. The diffusion coefficients are extrapolated from the normal diffusion regime characterized by:
\begin{eqnarray}
	\left\langle \Delta^{2} r\right\rangle = 6Dt, \qquad  t > \tau
 \end{eqnarray}
where $ \tau $ is the crossover time to the normal diffusion regime.\\
The Green-Kubo (GK) approach \cite{tovey2023mdsuite} that takes the autocorrelation with respect to the ionic current in the electrolyte $\mathbf{J}$ is used to calculate the ionic conductivity $\sigma$:
\begin{equation}
\label{eq:1}
 \sigma= \frac{V}{k_{B} T} \int_{0}^{\infty} dt \langle \mathbf{J}(t) \cdot \mathbf{J}(0) \rangle
 \end{equation}
 with,
 \begin{equation}
\label{eq:1q}
 \mathbf{J}(t) = q \sum_{i}^{N} z_{i}\mathbf{v}_{i}(t)
 \end{equation}
 Here $q$ is the elementary charge, $z_i$ represents the charge number (valence) of ion $i$, $\mathbf{v}_{i}$ is the velocity of ion $i$, $T$ is temperature, $k_B$ is the Boltzmann constant, $V$ is the volume of the simulation box, and $N$ is the number of ions. 

\begin{table*}[h!] 
  \centering 
  \caption{\normalsize Transport Properties of [EMIM-BF$_4$:DMSO (2:3) + 1M LiTFSI + ZnI$_2$ + CO$_2$] electrolyte from Molecular Dynamics simulation.}
  \label{Table5}
  \begin{tabular*}{\textwidth}{@{\extracolsep{\fill}}llll}
    \hline
    Parameter & Value & Unit & Symbol \\ 
    \hline
    Diffusion coefficient of Li$^{+}$ & 3.17 & 10$^{-9}$ m$^{2}$s$^{-1}$ & D$_{Li^{+}}$ \\ 
    Ionic conductivity of the electrolyte & 6.77 (7.4)$^{a}$ & mS cm$^{-1}$ & $\sigma$ \\ 
    Transference number of Li$^{+}$ & 0.91 (0.88)$^{a}$ & --- & $t_{+}$ \\ 
    Diffusion coefficient of CO$_{2}$ & 1.2 (1.16)$^{b}$ (1.14)$^{c}$ & 10$^{-9}$ m$^{2}$s$^{-1}$ & D$_{CO_{2}}$ \\ 
    \hline
  \end{tabular*}
  
  \vspace{0.5em} 
  \footnotesize 
  $^{a}$ Results for 1M LiTFSI/DMSO electrolyte \cite{tplus}\\
  $^{b}$ SEGWE Equation \cite{CO2DIFF}\\
  $^{c}$ Einstein equation \cite{CO2DIFF}\\
\end{table*}

The estimated diffusion coefficient of CO$_{2}$ in DMSO at 25 $\degree$C is approximately 1.16 $\times$ 10$^{-9}$ m$^2$/s using the Stokes-Einstein-Gierer-Wirtz (SEGWE) relation and 1.14 $\times$ 10$^{-9}$ m$^2$/s using the Einstein equation \cite{CO2DIFF}. These values are consistent with our MD calculation result of 1.2 $\times$ 10$^{-9}$ m$^2$/s presented in Table \ref{Table5}, further validating our findings.\\
The calculated diffusion coefficients, transference number, and ionic conductivity are directly incorporated into the FEA component of our multiscale modeling framework. These atomistically derived transport properties serve as critical input parameters for the FEA, enabling us to accurately simulate the intercalation and transport of CO$_{2}$ and ions during battery charge/discharge cycling at the cell level.

\subsubsection{Li$^+$ solvation structure from large-scale MD simulation}
The structural analysis was conducted utilizing the Radial Distribution Function (RDF), g(r), and Coordination Number (Nc). The RDF quantifies the spatial distribution of particles relative to a reference particle, providing insights into the local structural organization that affects the transport properties in the electrolyte.\\
The RDF is mathematically expressed as:
\begin{equation}
g_{\alpha \beta}(r)={\displaystyle \frac{\langle \rho_\beta(r) \rangle}{\langle\rho_\beta\rangle_{local}}}={\displaystyle \frac{1}{\langle\rho_\beta\rangle_{local}}}{\displaystyle \frac{1}{N_{\alpha}}}
                \sum_{i \in \alpha}^{N_{\alpha}} \sum_{j \in \beta}^{N_{\beta}}
                {\displaystyle \frac{\delta( r_{ij} - r )}{4 \pi r^2}}         
\end{equation}

with $\langle\rho_B(r)\rangle$ the particle density of type $\beta$ at a distance $r$ around particles $\alpha$, and $\langle\rho_B\rangle_{local}$ the particle density of type $\beta$ averaged over all spheres around particles $\alpha$ with radius $r_{max}$. Here, we chose $r_{max}$=12\AA.

The primary peak in g(r) corresponds to the nearest neighbors, indicating the most probable interparticle distance and interaction strength. This relationship is described by $g(r) = e^{(-\beta U(r))}$, where $\beta = 1/(k_B T)$, and $U(r)$ is the interaction potential. Subsequent peaks in the RDF elucidate higher-order neighbor relationships, offering insights into medium-range order, coordination numbers, and packing efficiency. In isotropic, homogeneous systems, g(r) asymptotically approaches unity at large distances, signifying a uniform particle distribution.

The coordination number, N$_c$, is defined as the number of neighboring particles within a specified cutoff distance of the pair interaction potential from a central particle. It is mathematically represented as:
\begin{equation}
    N_c = 2\pi \int_0^{r_c} n_r r dr = 2\pi n_b \int_0^{r_c} g(r) r dr
\end{equation}

where $r_c$ denotes the distance at which the interparticle interaction potential approaches zero, $n_b$ represents the bulk density, and $n_r$ is the average number density of particles at a given distance r, related to the RDF by $n_r$ = $n_b$ g(r).

\begin{figure*}[!h]
	\begin{center}
		\begin{minipage}[t]{1\textwidth}
					\centering
			\includegraphics[width=1\textwidth]{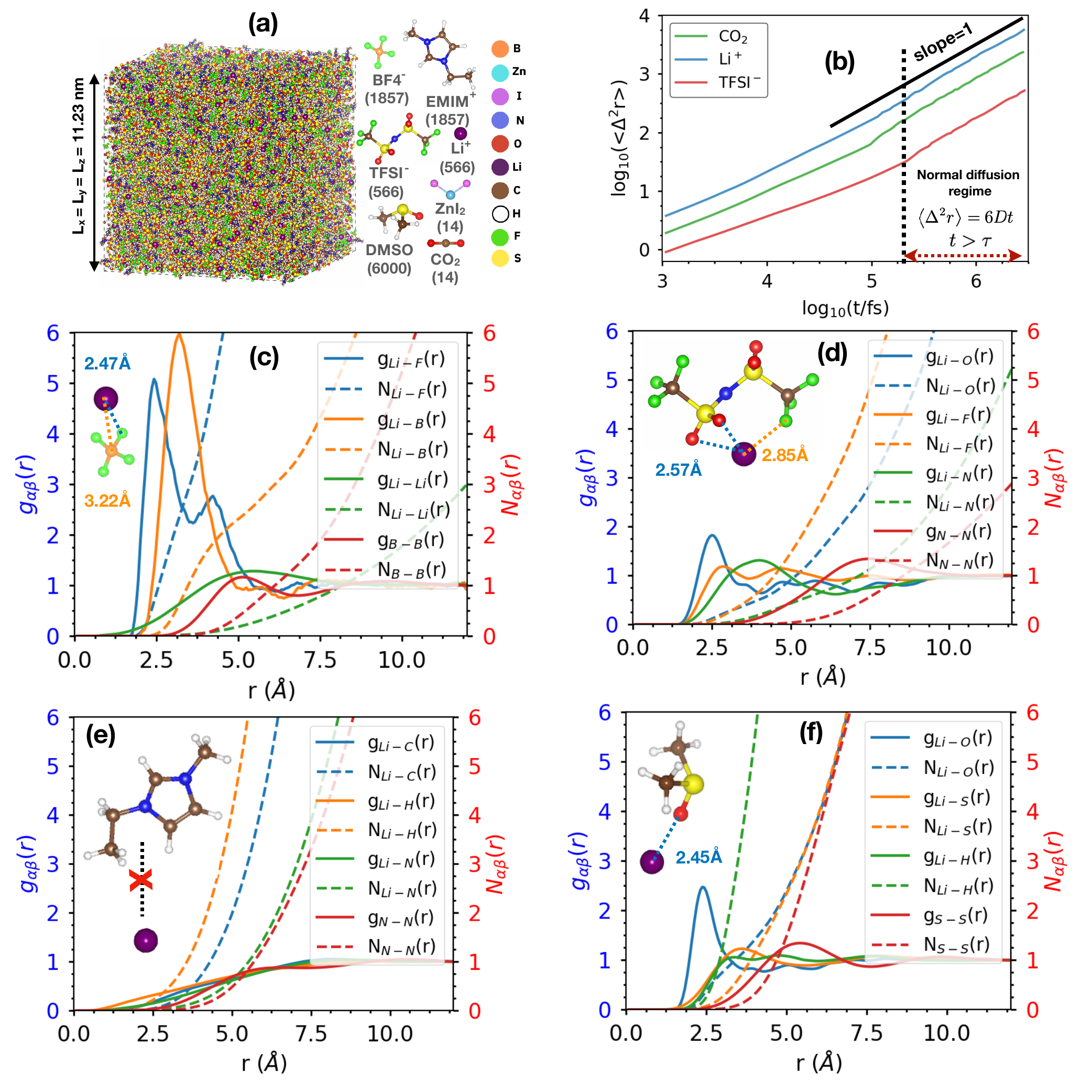}
			\caption{(a) A snapshot of the MD simulation of the liquid electrolyte, equilibrated at room temperature and ambient pressure (T = 300 K, P = 1 bar), is shown. The molecules comprising the electrolyte are depicted with their corresponding numbers in the simulation box. The atom colors used are also illustrated. (b) Mean-square-displacement of Li$^{+}$ , TFSI$^{-}$, and CO$_{2}$ whithin the electrolyte obtained from our MD simulation. (c), (d), (e), (f) Comparison of the average Radial Distribution Function (g(r)) and Coordination Number (N(r)) profiles for BF$_4$$^{-}$ (c), TFSI$^{-}$ (d), and EMIM$^{+}$ ions (e), as well as DMSO solvent (f), surrounding a single Li$^{+}$ ion. Additionally, the g(r) and N(r) between molecules of the same type are represented, which can be referred to as the g(r) and N(r) of the center-of-mass atoms of the molecules.}
            \label{Figure5} 
		\end{minipage}
	\end{center}
\end{figure*}

The coordination number as a function of distance, N(r), is expressed as:
\begin{equation}
N(r) = 2\pi n_b \int_0^r g(r') r' dr'
\end{equation}
where r represents an arbitrary distance from a reference particle.\\
We present the average Radial Distribution Function ($g(r)$) and Coordination Number ($N(r)$) profiles for BF$_4^-$, TFSI$^-$, and EMIM$^+$ ions, as well as DMSO solvent, surrounding a single Li$^+$ ion. Additionally, we analyze the $g(r)$ and $N(r)$ between molecules of the same type, represented by their center-of-mass atoms. Li$^+$ ions preferentially coordinate with molecules containing highly electronegative atoms, i.e., atoms with a negative partial charge. The flat $g(r)$ curves in Figure \ref{Figure5} (e) indicate a weak interaction between Li$^+$ and EMIM$^+$. Consequently, Li$^+$ primarily coordinates with BF$_4^-$, TFSI$^-$, and DMSO. According to Figure \ref{Figure5} (c), (d), and (f), Li$^+$ ions interact strongly with BF$_4^-$, TFSI$^-$, and DMSO. The most probable distances between Li$^+$ and the constituent atoms are: BF$_4^-$ (Li$^+$-F $\approx$ 2.47 \AA, Li$^+$-B $\approx$ 3.22 \AA), TFSI$^-$ (Li$^+$-O $\approx$ 2.57 \AA, Li$^+$-F $\approx$ 2.85 \AA), and DMSO (Li$^+$-O $\approx$ 2.45 \AA). The Li$^+$ cations in the electrolyte are more solvated by BF$_4^-$ and DMSO than TFSI$^-$, as evidenced by the higher peaks in the $g(r)$ and the corresponding coordination number curves of Li$^+$-BF$_4^-$. BF$_4^-$ and DMSO are closer to the Li$^+$ ions than TFSI$^-$ ions, playing a crucial role in the dissociation of LiTFSI salt. However, TFSI$^-$ can also access the first coordination shell of Li$^+$ due to strong interactions between its O and F atoms and Li$^+$. The presence of both BF$_4^-$ and TFSI$^-$ facilitates favorable coordination between Li$^+$ and F, potentially leading to the formation of a Li-F solid electrolyte interphase (SEI) that protects the Li-metal anode. Interactions between molecules of the same type are weak due to electrostatic repulsion, resulting in well-dispersed molecules throughout the simulated electrolyte. The average first shell coordination number values for a tagged Li$^+$ cation were determined using a threshold distance of 3 \AA, based on the range of the first $g(r)$ peaks. Key findings from the coordination number analysis include: Li$^+$-Li$^+$ $N(r=3\text{\AA}) \approx 0$) and $N(r=5\text{\AA}) \approx 0.25$), Li$^+$-BF$_4^-$ ($N_{\text{Li-F}}(r=3\text{\AA}) \approx 2$, $N_{\text{Li-B}}(r=3\text{\AA}) \approx 0.25$), Li$^+$-DMSO ($N_{\text{Li-O}}(r=3\text{\AA}) \approx 0.8$), and Li$^+$-TFSI$^-$ ($N_{\text{Li-O}}(r=3\text{\AA}) \approx 0.25$ and $N_{\text{Li-F}}(r=3\text{\AA}) \approx 0.25$). \\

The high ionic conductivity (Table \ref{Table5}) of the simulated electrolyte is derived from several key factors revealed by the solvation structure and interaction analyses (Figure \ref{Figure5}). Li$^+$ cations are preferentially solvated by BF$_4^-$ and DMSO, promoting LiTFSI salt dissociation and increasing free charge carriers. Weak Li$^+$-TFSI$^-$ interactions facilitate easier Li$^+$ transport. The presence of BF$_4^-$ and TFSI$^-$ creates Li$^+$ conducting channels through favorable Li$^+$-F coordination. Well-dispersed ionic species due to weak same-type molecular interactions enhance efficient ion transport. Optimal coordination numbers (e.g., $N_{\text{Li-F}}(r=3\text{Å}) \approx 2$ for BF$_4^-$, $N_{\text{Li-O}}(r=3\text{Å}) \approx 0.8$ for DMSO) allow efficient Li$^+$ transport while maintaining sufficient solvation. Low Li$^+$-Li$^+$ coordination ($N(r=3\text{Å}) \approx 0$) minimizes clustering, promoting ion mobility. Multiple solvating species create a dynamic environment, facilitating rapid solvating molecule exchange and enhancing Li$^+$ mobility. The preferential solvation of Li$^+$ by BF$_4^-$ and DMSO allows for more free movement compared to the larger TFSI$^-$ anions, resulting in a higher proportion of the ionic current carried by Li$^+$ and thus increasing the Li$^+$ transference number.

\subsection{Cell performance from Finite Element Analysis}
\begin{figure*}[!h]
	\begin{center}
		\begin{minipage}[t]{1\textwidth}
					\centering
			\includegraphics[width=1\textwidth]{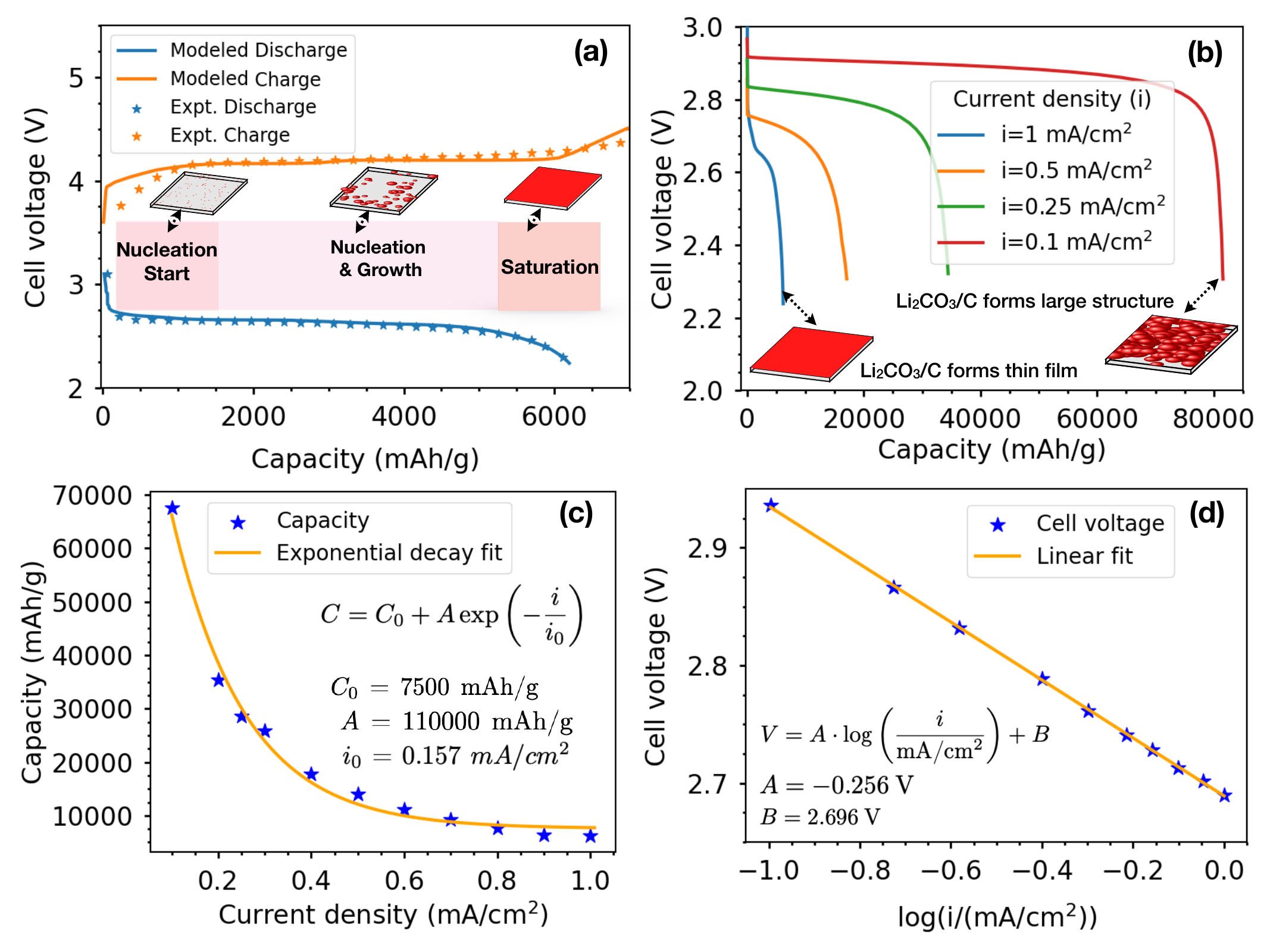}
            \caption{(a) Deep discharge/charge curve of the Li-CO$_{2}$ battery cell operating at 1 mA/cm$^{2}$, obtained from our multiscale model and experiment. (b) Deep discharge curves of the Li-CO$_2$ battery cell operating at different applied current densities of 1 mA/cm$^{2}$, 0.5 mA/cm$^{2}$, 0.25 mA/cm$^{2}$, and 0.1 mA/cm$^{2}$. The insets in (a) and (b) show a schematic representation of the Li$_{2}$CO$_{3}$/C formation process (red color) on the cathode's CO$_{2}$ feed side surface (gray color). (c) Capacity as a function of discharge current density. (d) Cell potential at the start of the discharge voltage plateau (Li$_{2}$CO$_{3}$/C nucleation and growth) with respect to current density.}
            \label{Figure6} 
		\end{minipage}
	\end{center}
\end{figure*}

The simulated voltage-capacity curves of the Li-CO$_2$ battery cell during charge/discharge cycles are shown in Figure ~\ref{Figure6} (a). These curves were generated through Finite Element Analysis (FEA) using input parameters derived from atomistic simulations (Figure ~\ref{Figure3}, Table ~\ref{Table5}) and additional material properties detailed in Tables \ref{Table1},  \ref{Table2},  and \ref{Table3}. The FEA results were validated against experimental data collected at a current density of 1 mAh/cm$^2$, as described in the Supporting Information (SI). A strong correlation between simulated and experimental voltages is observed. During discharge, the voltage profile exhibits three distinct regions: (1) an initial sharp voltage drop due to kinetic overpotential ($\phi_{1} - \phi_{2}-E^0$) \cite{viswanathan2013li}, (2) a linear decline attributed to resistive losses ($\phi_{\text{film}}$) across the growing Li$_2$CO$_3$/C film (average plateau potential: 2.65 V), (3) a "sudden death" characterized by rapid voltage collapse after 5000 mAh/g. The latter two phenomena, previously described for Li-O$_2$ batteries as charge-transport limitations through Li$_2$O$_2$ films \cite{viswanathan2011electrical}, exhibit analogous behavior in Li-CO$_2$ systems. During charging, the cell maintains a stable potential of approximately 4.2 V until reaching 6000 mAh/g, followed by a steep voltage rise. The system achieves a specific capacity of 6200 mAh/g (0.48 mAh/cm$^2$).\\
Figure \ref{Figure6} (b) illustrates the impact of applied current density of 1 mA/cm$^{2}$, 0.5 mA/cm$^{2}$, 0.25 mA/cm$^{2}$, and 0.1 mA/cm$^{2}$ on the simulated discharge curves of Li-CO$_{2}$ battery cell and Figure S1 shows the same curves for current densities of 0.2 mA/cm$^{2}$, 0.3 mA/cm$^{2}$, 0.4 mA/cm$^{2}$, 0.6 mA/cm$^{2}$, 0.7 mA/cm$^{2}$, 0.8 mA/cm$^{2}$ and 0.9 mA/cm$^{2}$. The discharge capacity exhibits a substantial reduction, decreasing from 81,570 mAh/g at a low current density of 0.1 mA/cm$^{2}$ to 6,200 mAh/g at the high current density of 1 mA/cm$^{2}$. This observation aligns with prior experimental results, indicating that current density has a significant influence on cell capacity \cite{li2019drawing}. Furthermore, the discharge voltage plateau decreases as the current density increases (Figure \ref{Figure6} (b)). The capacity decrease at elevated discharge rates can be attributed to limitations in CO$_{2}$ transport through the cathode, which becomes saturated with electrolyte and unable to sustain the electrochemical reaction. Consequently, as the discharge rate increases, CO$_{2}$ reduction is confined to a localized area near the cathode-current collector interface. Furthermore, the rapid reduction in porosity due to Li$_{2}$CO$_{3}$ and C deposition on the active area's surface further impedes CO$_{2}$ transport into the cell, preventing complete utilization of the electrode's porosity.

\begin{figure*}[!h]
	\begin{center}
		\begin{minipage}[t]{1\textwidth}
					\centering
			\includegraphics[width=1\textwidth]{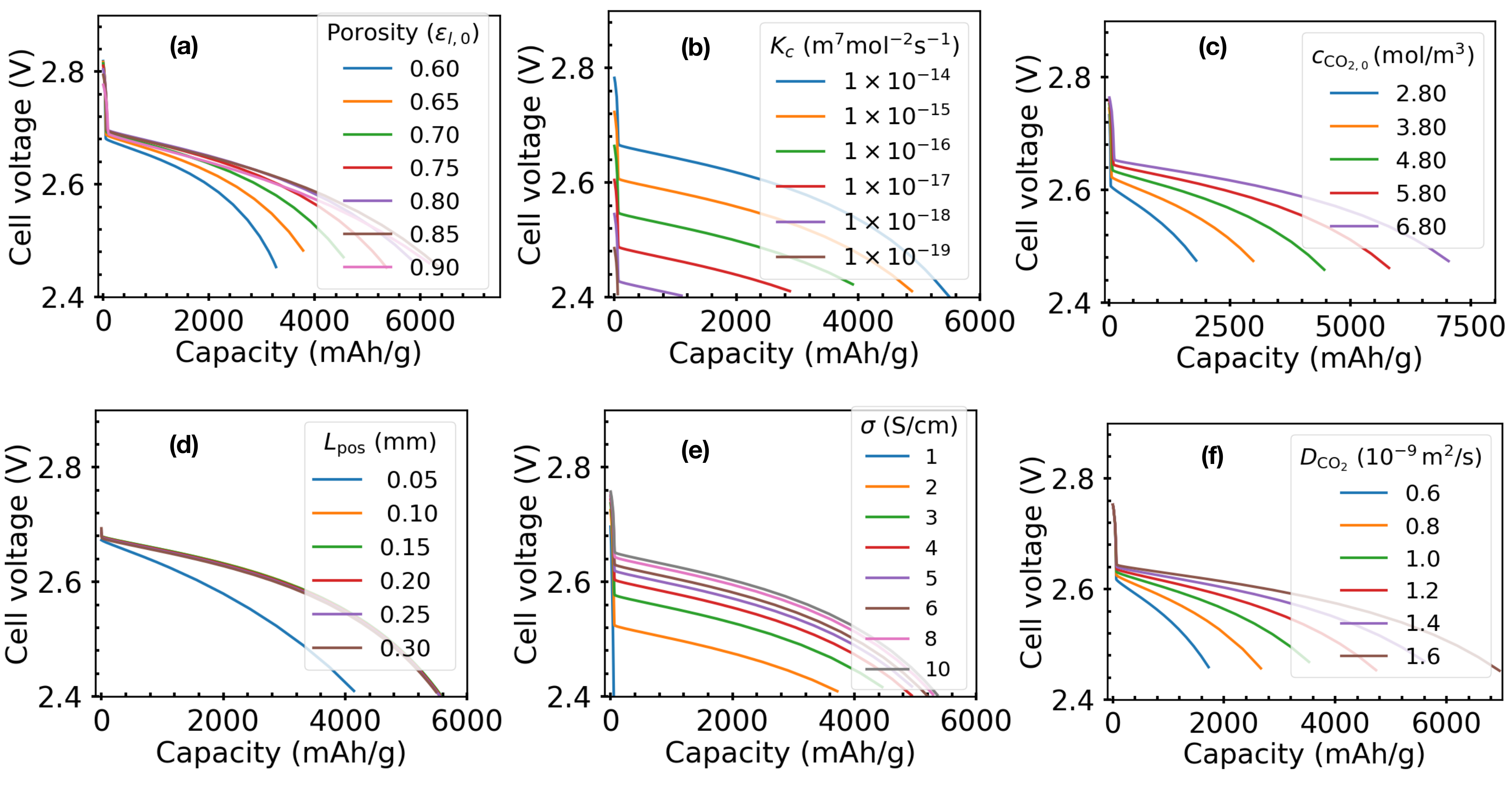}
            \caption{Impact of key parameters on voltage–capacity profiles: (a) Initial porosity ($\varepsilon_{l,0}$), (b) cathodic reaction rate constant ($K_c$), (c) initial c$_{CO_{2, 0}}$ concentration, (d) cathode thickness ($L_{\text{pos}}$), (e) ionic conductivity ($\sigma$), and (f) CO$_2$ diffusion coefficient ($D_{\text{CO}_2}$). The current density was held constant at 1 mA/cm$^{2}$; all other parameters are used as provided in Tables \ref{Table1}, \ref{Table2}, and \ref{Table3}.}
            \label{Figure7} 
		\end{minipage}
	\end{center}
\end{figure*}

\begin{figure*}[!h]
	\begin{center}
		\begin{minipage}[t]{1\textwidth}
					\centering
			\includegraphics[width=1\textwidth]{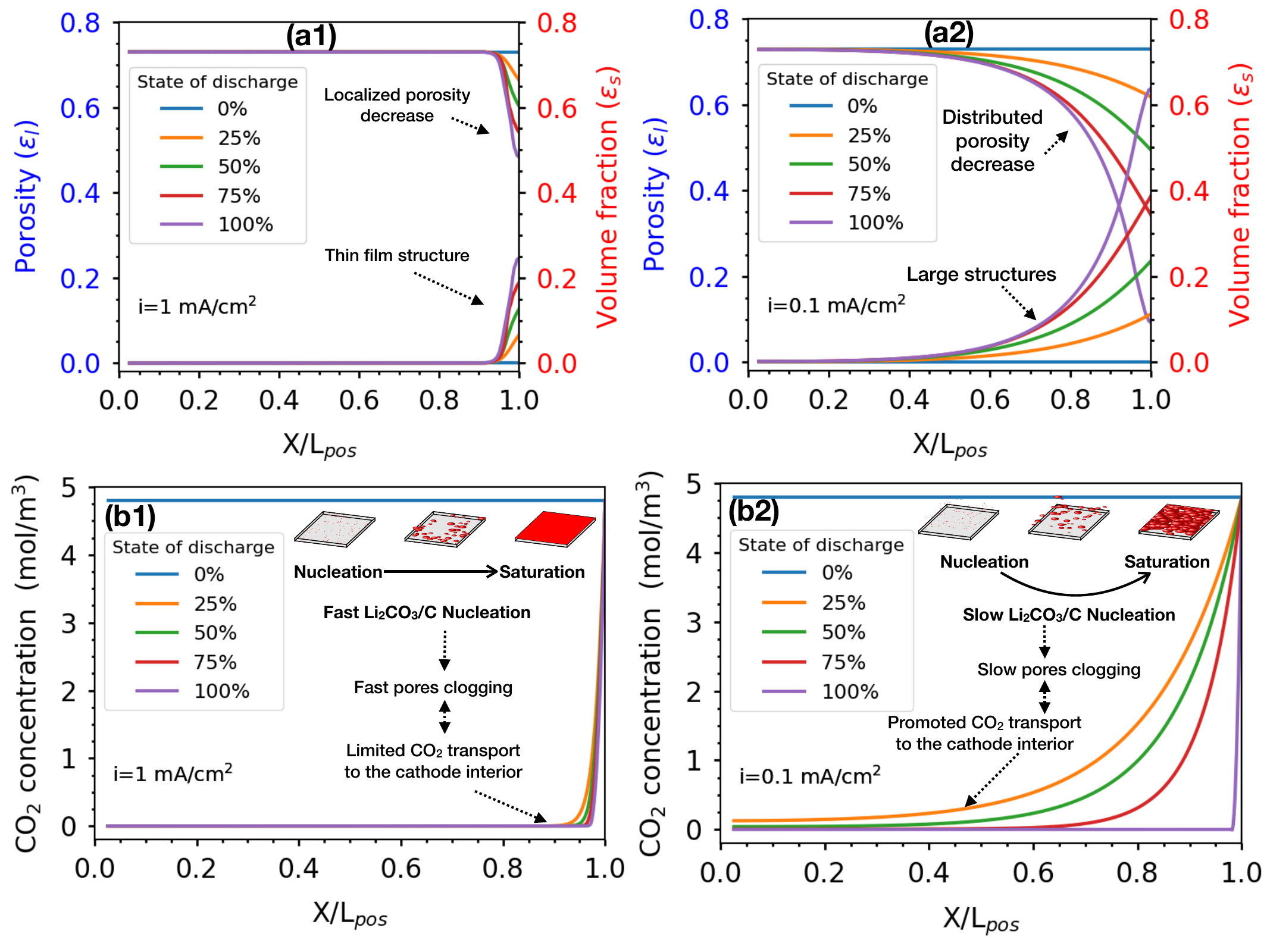}
            \caption{The changes in porosity and volume fraction of deposited species throughout the discharge process are shown at different current densities ((a1) 1 mA/cm$^{2}$ and (a2) 0.1 mA/cm$^{2}$) in relation to the state of discharge. The local CO$_{2}$ concentration within the Li-CO$_{2}$ cathode during discharge was analyzed at rates of 0.1 mA/cm$^{2}$ (b1) and 1 mA/cm$^{2}$ (b2), corresponding to different discharge states (0\% indicates a fully charged battery and 100\% indicate a fully discharged battery). The distance (X) is normalized by the cathode thickness (L$_{pos}$). The insets in (b1) and (b2) show a schematic representation of the different nucleation processes leading to different Li$_{2}$CO$_{3}$/C formation structures depending on current density.}
            \label{Figure8} 
		\end{minipage}
	\end{center}
\end{figure*}
 Figure \ref{Figure6} (c) demonstrates that the capacity of the studied Li-CO$_2$ battery cell decreases exponentially with increasing current density, which can be described by the following equation \cite{griffith2015correlating}:
 \begin{equation} C = C_0 + A \exp\left(-\frac{i}{i_0}\right) \end{equation}\\
 where:  $C_0$ = 7500 mAh/g is the baseline capacity, $A$ = 110000 mAh/g is a scaling factor, $i_0$ = 0.157 $mA/cm^2$ is a characteristic current density. At low current density (i $\ll$ 0.157 $mA/cm^2$), the capacity $C$ tends towards its maximum value of 117500 mAh/g. This is the sum of the baseline capacity ($C_0$) and the additional capacity represented by $A$. As the current density $i$ increases, there is a rapid decrease in the additional capacity provided by $A$, resulting in a decrease in the total capacity $C$. The parameter $i_0$ determines the rate at which the capacity decreases with increasing current density. A smaller $i_0$ would mean a faster decrease in capacity for a given increase in current density, while a larger $i_0$ would indicate a slower decrease.  As the current density increases further, the capacity approaches the baseline value $C_0$ of 7500 mAh/g. This suggests that at high current densities, the battery's capacity is primarily limited to its baseline capacity, as the contribution from the exponential term becomes negligible. The exponential decrease in capacity with increasing current density highlights a trade-off between power output and energy capacity \cite{quarti2023trade}. At higher current densities (i $\gg$ 0.157 $mA/cm^2$), the battery can deliver more power, but at the cost of reduced capacity. This is a critical consideration in applications where both high power and high capacity are required. The connection between cell voltage ($V$) and current density ($i$) during the discharge of the studied Li-CO$_2$ battery cell in Figure \ref{Figure6} (d) is represented by the formula \cite{griffith2015correlating, xiao2023unveiling}:  
 \begin{equation} 
 V = A \cdot \log\left(\frac{i}{\text{mA/cm$^{2}$}}\right) + B
 \end{equation}\\
where $A = -0.256$ V and $B = 2.696$ V. The parameter $A$  represents the slope of the voltage change with the logarithm of the current density/(mA/cm$^{2}$), while  $B$  represents the voltage at a reference current density of 1 mA/cm$^2$. This formula indicates that as the current density increases, the cell voltage decreases logarithmically, a common characteristic in electrochemical cells due to increased polarization and decreased voltage at higher current densities \cite{griffith2015correlating, xiao2023unveiling}. \\
Figure \ref{Figure7} summarizes the influence of key parameters on the voltage–capacity profiles of Li-CO$_2$ batteries at a current density of 1 mA/cm$^{2}$. To investigate the influence of key parameters on the voltage-capacity profiles (Figure 7), we performed a parameter sensitivity analysis using FEA to solve the model equations described in Section 2.4. The base case parameters, as provided in Tables 1, 2, and 3, were used as the reference. Each parameter was varied over a specified range while keeping the others at their base values. Specifically, the initial porosity ($\varepsilon_{l,0}$) was varied from 0.6 to 0.9 in steps of 0.05, the cathodic reaction rate constant ($K_c$) from 1 $\times$ 10$^{-19}$ to 1 $\times$ 10$^{-14}$ m$^7$/s/mol$^2$ (logarithmically spaced), the initial CO$_{2}$ concentration (c$_{CO_{2, 0}}$) from 2.8 to 6.8 mol/m$^{3}$, the cathode thickness ($L_{\text{pos}}$) from 0.05 to 0.3 mm, the ionic conductivity ($\sigma$) from 1 to 10 mS/cm, and the CO$_{2}$ diffusion coefficient ($D_{\text{CO}_2}$) from 0.6 $\times$ 10$^{-9}$ m$^{2}$/s to 1.6 $\times$ 10$^{-9}$ m$^{2}$/s. For each set of parameters, the discharge process was simulated at a constant current density of 1 mA/cm$^{2}$. As shown in Figure \ref{Figure7} (a), increasing the initial porosity ($\varepsilon{l,0}$) leads to higher battery capacity by expanding the available surface area and enhancing CO$_2$ accessibility, thereby promoting more efficient electrochemical reactions. Figure \ref{Figure7} (b) highlights the critical role of the cathodic reaction rate constant ($K_c$); the voltage–capacity profile exhibits high sensitivity to variations in $K_c$, emphasizing the necessity of effective catalytic materials for optimal battery performance. In Figure \ref{Figure7} (c), it is evident that higher initial CO$_2$ concentration ($c{\text{CO}{2,0}}$) results in increased capacity and a more stable voltage plateau, attributed to improved reactant availability at the cathode. Figure \ref{Figure7} (d) demonstrates that increasing the cathode thickness ($L_{\text{pos}}$) from 0.05 mm to 0.1 mm substantially enhances capacity due to the greater active volume; however, further increases beyond 0.1 mm do not provide additional capacity gains, likely due to diffusion limitations. Figure \ref{Figure7} (e) shows that electrolyte ionic conductivity ($\sigma$) is a critical factor, with values above 2 mS/cm required to achieve high capacities (>4000 mAh/g) and improved voltage stability. Figure \ref{Figure7} (f) indicates that a higher CO$_2$ diffusion coefficient ($D{\text{CO}_2}$) further enhances both capacity and voltage plateau by facilitating more efficient CO$_2$ transport to active sites. Notably, the same trends are observed at lower current densities; Figure S2 depicts these effects at a current density of 0.4 mA/cm$^2$. Taken together, these results underscore the importance of simultaneously optimizing electrode structure, electrolyte properties, and catalytic activity to maximize the performance of Li-CO$_2$ batteries.\\
Figure \ref{Figure8} (a1) and (a2) illustrate how porosity ($\varepsilon_l$) and volume fraction of Li$_2$CO$_3$/C deposited species ($\varepsilon_s$) change over space and time with current densities of i=1 mA/cm$^2$ and i=0.1 mA/cm$^2$, respectively. Figure S3 presents a comparison of these parameters at i=0.1, i=0.25, i=0.5, and i=1 mA/cm$^2$. For high current density (i=1 mA/cm$^2$), porosity decreases gradually, primarily near the CO$_{2}$ feed side of the cathode (X/L${pos}$=1), from its initial value of $\varepsilon_l=0.73$ to 0.48 at the end of discharge. This decrease is due to the formation of a thin film (flake-like) of Li$_2$CO$_3$/C \cite{peng2023influence}, represented by the gradual increase of $\varepsilon_s$ from 0 to 0.25 near the cathode feed side. This film formation leads to electrode clogging, impeding efficient CO$_2$ diffusion into the cathode. For low current density (i=0.1 mA/cm$^2$), Li$_2$CO$_3$/C forms larger structures \cite{peng2023influence}, evidenced by the gradual increase of $\varepsilon_s$ across a wider region of the cathode from 0 to 0.64. This results in a distributed porosity decrease from 0.73 to 0.09, initially allowing more CO$_{2}$ transport into the cathode. However, as discharge progresses, the substantial porosity decrease hinders CO$_{2}$ diffusion even at low current densities. Figure \ref{Figure8} (b1) and (b2) present the CO$_{2}$ concentration profiles within the cell during discharge at i=1 mA/cm$^2$ and i=0.1 mA/cm$^2$, respectively. Figure S4 compares concentration profiles at various current densities. At i=1 mA/cm$^2$, the CO$_{2}$ concentration ($c_{CO_2}$) decreases sharply from its initial value of 4.8 mol/m$^3$ to 0 mol/m$^3$ in the porous cathode, indicating strongly impeded CO$_{2}$ diffusion. At i=0.1 mA/cm$^2$, the decrease is more gradual. The impediment to CO$_{2}$ transport, which is inversely proportional to the current density, results from the accumulation of Li$_2$CO$_3$ and C on the active cathode surface. This accumulation reduces pore availability for the electrolyte, increases the amount of insulating species, and obstructs pores, collectively restricting CO$_{2}$ transport and limiting cell capacity (Figure \ref{Figure6} (c)). The nucleation and growth of Li$_2$CO$_3$ significantly impact performance, as described in Figure \ref{Figure4} and the detailed FEA model in section \ref{section2.4}. Current density plays a crucial role in this process. At high current densities (1 mA/cm$^2$), rapid nucleation near the CO$_2$ feed side forms a thin Li$_2$CO$_3$/C film (Figure \ref{Figure8} (a1), (b1)). In contrast, low current densities (0.1 mA/cm$^2$) promote slower growth, resulting in larger, more distributed Li$_2$CO$_3$ structures and a gradual decrease in porosity (Figure \ref{Figure8} (a2), (b2)).\\
To address these limitations, an ideal cathode should feature: a large surface area for reaction sites and product deposition, large pores to facilitate active species transport and prevent clogging, high porosity to accommodate discharge product growth, and high electrical conductivity and stability. Such a cathode design could mitigate the observed limitations and potentially enhance the overall performance of Li-CO$_{2}$ batteries \cite{ss2023binder}. Finally, we note that direct experimental validation is essential to confirm our model’s mechanistic predictions. To rigorously assess accuracy and applicability, high-resolution imaging techniques such as transmission electron microscopy (TEM) and scanning electron microscopy (SEM) should be used to verify predicted changes in Li$_2$CO$_3$/C morphology, cathode surface features, and Li$_2$CO$_3$/C aggregation states. Additionally, time-resolved or ex-situ X-ray diffraction (XRD) can track the evolution of Li$_2$CO$_3$/C crystal structure and phase composition, while gas adsorption measurements, including Brunauer-Emmett-Teller (BET) and Barrett-Joyner-Halenda (BJH) analysis, can provide a quantitative assessment of cathode porosity and specific surface area. Importantly, a multiscale experimental approach is also needed to capture material behavior across different length and time scales, ensuring that both nanoscale mechanisms and macroscopic properties are accurately linked and described. These targeted and multiscale experiments are crucial for establishing the model’s predictive accuracy and real-world relevance.
\section{Conclusion} \label{Conclusion}
Our interactive multiscale modeling framework, integrating FEA, MD, and DFT/AIMD, provides comprehensive insights into Li-CO$_2$ battery electrochemistry. This approach enables accurate prediction of cell performance by considering critical factors such as CO$_2$ transport, cathode porosity, and Li$_2$CO$_3$/C deposition.
DFT/AIMD calculations reveal the high electrical conductivity of the cathode catalyst, while MD simulations demonstrate favorable electrolyte properties. FEA results show an exponential decrease in discharge capacity with increasing current density-from 81,570 mAh/g at 0.1 mA/cm$^2$ to 6,200 mAh/g at 1 mA/cm$^2$-primarily due to CO$_2$ transport limitations and reduced cathode porosity. The morphology of Li$_2$CO$_3$/C deposits transitions from larger structures at lower current densities to thin films at higher current densities, significantly impacting performance.\\

Our multiscale approach bridges atomic-level phenomena and cell-level performance, offering unique insights into the spatiotemporal evolution of key battery parameters and reducing the need for extensive experimental prototyping. Future research should focus on enhancing CO$_2$ transport, improving discharge product decomposition, and optimizing cathode design with large surface areas, optimal pore sizes, and high porosity. This framework accelerates the development of high-performance Li-CO$_2$ batteries and sustainable energy storage technologies by providing insights into phenomena that are difficult to observe experimentally in real time. Moreover, the developed Li-CO$_2$ battery continuum model at the cell level can be extended to study battery packs and thermal runaway. This modeling framework can also be adapted to other battery chemistries or used to simulate degradation, cycling behavior, and thermal effects, thereby broadening its applicability.
\section*{CRediT Authorship Contribution Statement}
Mohammed Lemaalem, Selva C. Selvaraj, and Naveen K. Dandu: Conceptualization, Visualization, Software, Data Curation, Formal Analysis, Methodology, Writing – Original Draft, Writing – Review \& Editing. Ilias Papailias and Arash Namaeighasemi: Methodology, Data Curation, Formal Analysis, Writing – Review \& Editing. Larry A. Curtiss and Amin Salehi-Khojin: Supervision, Data Curation, Formal Analysis, Writing – Review \& Editing. Anh T. Ngo: Supervision, Conceptualization, Data Curation, Formal Analysis, Investigation, Project Administration, Writing – Review \& Editing.
\section*{Author Contributions}
This work was conceived by A. T. N.. M. L., S. C. S., and N. K. D., carried out the calculations, including first-principles DFT, AIMD results, large-scale MD simulations, and Finite Element Analysis. I. P., A. N., L. A. C., and A. S. K. were responsible for the experimental measurements. All authors contributed to the discussion of the results and the final manuscript.
\section*{Declaration of Competing Interest}
The authors declare that they have no known competing financial interests or personal relationships that could have appeared to influence the work reported in this paper.
\section*{Acknowledgements}
This work was supported by the Assistant Secretary for Energy Efficiency and Renewable Energy, Office of Vehicle Technologies of the US Department of Energy, through the Battery Materials Research (BMR) program. We gratefully acknowledge the computing resources provided on Bebop, a high-performance computing cluster operated by the Laboratory Computing Resource Center (LCRC) at Argonne National Laboratory.
\section*{\textbf{Appendix A. Supplementary material}}
Background on catalyst/electrolyte developments for Li-CO$_2$ battery; Description of the experimental setup; Computational details; Plots showing the voltage-capacity curve during Li-CO$_2$ battery discharge at various rates ranging from 0.2 mA/cm$^{2}$ to 0.9 mA/cm$^{2}$; Plots showing the effect of initial values of relevant variables on the voltage-capacity curve during Li-CO$_2$ battery discharge at a rate of 0.4 mA/cm$^{2}$; Plots illustrating changes in porosity and volume fraction of deposited species in the cathode throughout the discharge process at different current densities and discharge states; and Plots showing the change in CO$_2$ concentration within the cathode.
\section*{Data availability}
The data that support the plots within this paper and other findings of this study are available from the corresponding author upon reasonable request.

\bibliography{rsc}
\bibliographystyle{elsarticle-num}

\begin{figure*}[!h]
	\begin{center}
		\begin{minipage}[t]{1\textwidth}
        \captionsetup{labelformat=empty}
        \caption{\large\textbf{Graphical abstract}}
					\centering
			\includegraphics[width=0.8\textwidth]{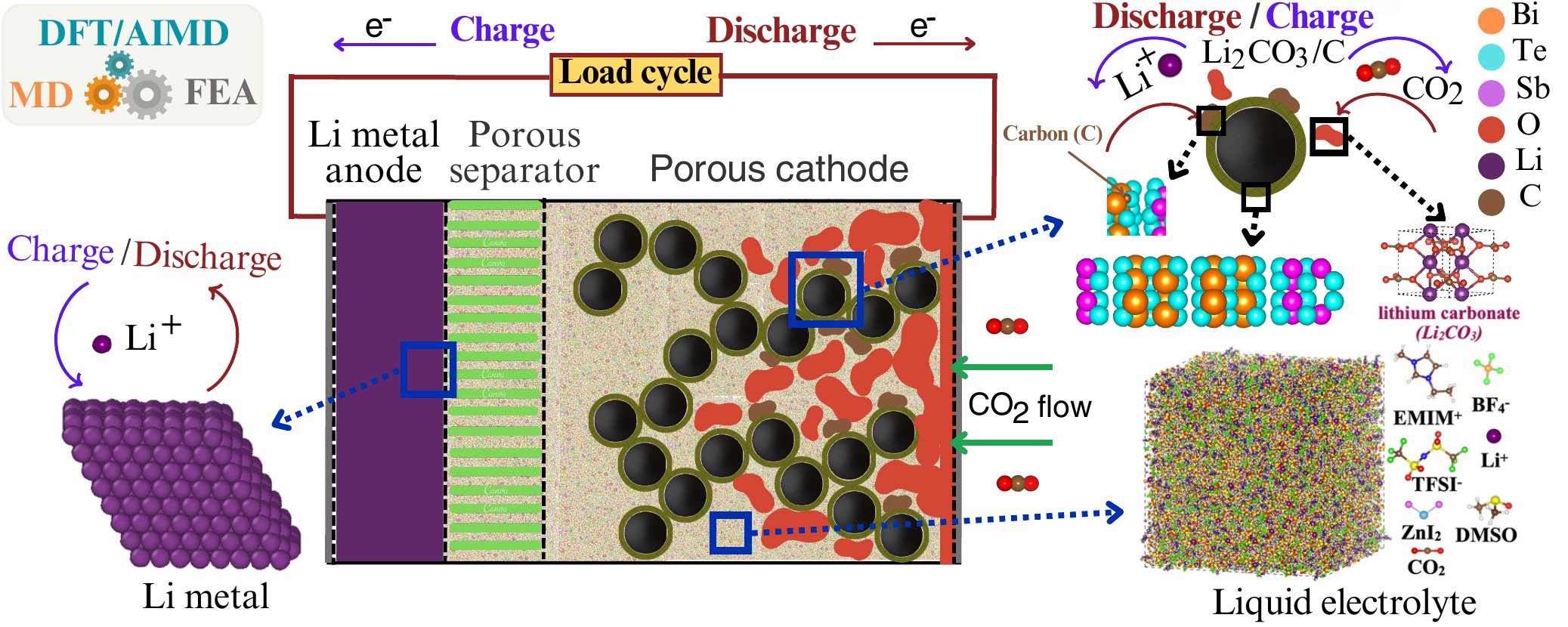}
             
		\end{minipage}
	\end{center}
\end{figure*}

\includepdf[pages=-]{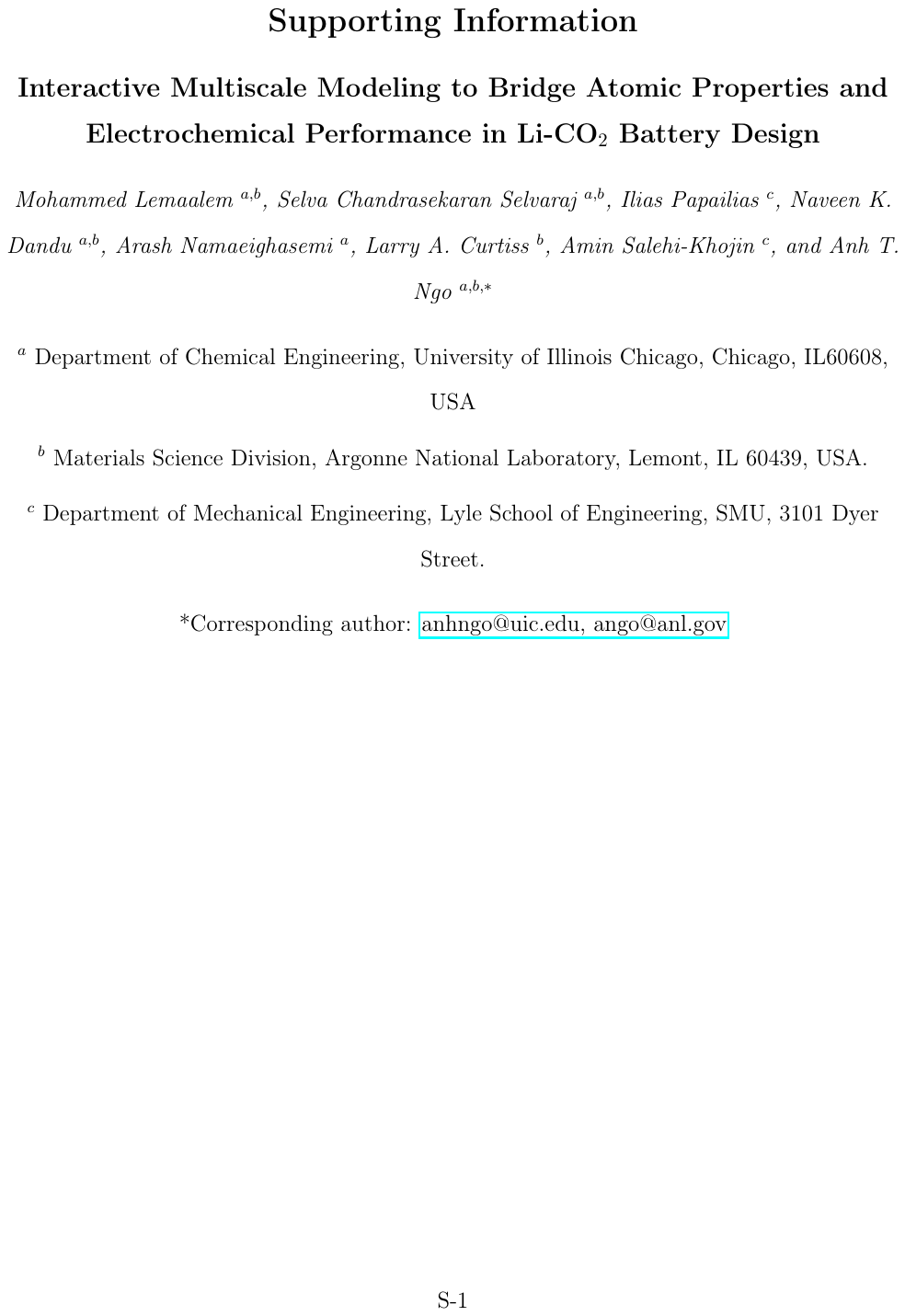}
\end{document}